\documentclass[]{aa}

\usepackage[utf8]{inputenc}
\usepackage{booktabs}
\usepackage{float}
\usepackage[caption = false]{subfig}
\usepackage{graphicx}
\usepackage{natbib}
\bibpunct{(}{)}{;}{a}{}{,} 
\usepackage{txfonts}
\usepackage{booktabs}
\usepackage{amsmath}
\usepackage{amsfonts}
\usepackage{amssymb}
\bibpunct{(}{)}{;}{a}{}{,}
\usepackage[english]{babel}

\usepackage[table]{xcolor}
\usepackage{longtable}

\usepackage{placeins}

\usepackage[normalem]{ulem}

\usepackage{url}
\usepackage{hyperref}
\hypersetup{
    colorlinks = true,
    linkcolor = blue,
    anchorcolor = blue,
    citecolor = blue,
    filecolor = blue,
    urlcolor = blue
    }

\newcommand{\orcid}[1]{\href{https://orcid.org/#1}{\includegraphics[width=10pt]{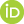}}}

\newcommand{\FeIIVelPeak}{34{,}800 \pm 2{,}300\, \mathrm{(random)} \pm 8{,}000 \, \mathrm{(systematic)}}
\newcommand{\FeIIVelPeakVELONLY}{34{,}800}

\newcommand{\Nimass}{0.23\pm0.03 }
\newcommand{\Mej}{3.2\pm0.8}
\newcommand{\EK}{(23.1\pm12.4)} 
\newcommand{\EKnoerr }{23.1}
\newcommand{\EKoverM}{(7.2\pm3.5)}

\newcommand{\TimeLmax}{8.8}

\newcommand{\Mejsyn}{3.5}
\newcommand{\EKsyn}{17}
\newcommand{\EoverMsyn}{4.9}
\newcommand{\kms}{\ensuremath{\mathrm{km\ s^{-1}}}}

\newsavebox\ltmcbox

\begin{document}

\title{The broad-lined type Ic supernova 2020lao experienced an energetic explosion with no central-engine signatures}

\titlerunning{The infant broad-linxe type~Ic  SN\,2020lao}

\authorrunning{Stritzinger et al.}

\author{ 
M. D. Stritzinger\inst{\ref{uni:aarhus}}\orcid{0000-0002-5571-1833}
\and 
T. J. Moriya\inst{\ref{uni:tokyo},\ref{uni:sokendai},\ref{uni:monash}}\orcid{0000-0003-1169-1954}
\and 
S. Bose\inst{\ref{uni:aarhus}}\orcid{0000-0003-3529-3854}
\and
P. A. Mazzali\inst{\ref{uni:LJM}}\orcid{0000-0001-6876-8284}
\and 
P.~Lundqvist\inst{\ref{uni:stockholm}\orcid{0000-0003-0065-2933}}
\and 
E. Karamehmetoglu\inst{\ref{uni:aarhus}}\orcid{0000-0001-6209-838X} 
\and 
L. S. Arndt\inst{\ref{uni:aarhus}}\orcid{0009-0005-1809-3617}
\and 
C. Ashall\inst{\ref{uni:hawaii}}\orcid{0000-0002-5221-7557}
\and
L.~Galbany\inst{\ref{uni:spain1},\ref{uni:spain2}}\orcid{0000-0002-1296-6887}
\and 
W. B. Hoogendam\inst{\ref{uni:hawaii}}\orcid{0000-0003-3953-9532}
\and
E. Baron\inst{\ref{uni:baron1}}\orcid{0000-0001-5393-1608}
\and 
J.~M.~DerKacy\inst{\ref{baltimore}}\orcid{0000-0002-7566-6080}
\and
N. Elias-Rosa\inst{\ref{uni:padua}}\orcid{0000-0002-1381-9125} 
\and 
E. Y. Hsiao\inst{\ref{uni:FSU}}\orcid{0000-0003-1039-2928}
\and 
P. Höflich\inst{\ref{uni:FSU}}\orcid{0000-0002-4338-6586}
\and 
E. Pian\inst{\ref{uni:pisa}}\orcid{0000-0001-8646-4858}
\and 
E. A. M. Jensen\inst{\ref{uni:aarhus}}\orcid{0000-0003-3197-3430}
\and 
S. Moran\inst{\ref{uni:UK}}\orcid{https://orcid.org/0000-0001-5221-0243}
\and 
A. Pastorello\inst{\ref{uni:padua}}\orcid{0000-0002-7259-4624}
\and 
M. Shahbandeh\inst{\ref{baltimore}}\orcid{0000-0002-9301-5302} 
\and 
G. Valerin\inst{\ref{uni:padua}}\orcid{0000-0002-3334-4585}
}

\institute{Department of Physics and Astronomy, Aarhus University, Ny Munkegade 120, DK-8000 Aarhus C, Denmark\label{uni:aarhus}
\email{max@phys.au.dk}
\and 
National Astronomical Observatory of Japan, National Institutes of Natural Sciences, 2-21-1 Osawa, Mitaka, Tokyo, 181-8588, Japan\label{uni:tokyo}
\and
Graduate Institute for Advanced Studies, SOKENDAI, 2-21-1 Osawa, Mitaka, Tokyo 181-8588, Japan\label{uni:sokendai}
\and
School of Physics and Astronomy, Monash University, Clayton, VIC 3800, Australia\label{uni:monash}
\and
Astrophysics Research Institute, Liverpool John Moores University, IC2, Liverpool Science Park, 146 Brownlow Hill, Liverpool L3 5RF, UK\label{uni:LJM}
\and
The Oskar Klein Centre, Department of Astronomy, Stockholm University, AlbaNova, 10691 Stockholm, Sweden\label{uni:stockholm}
\and 
Institute for Astronomy, University of Hawai'i, 2680 Woodlawn Drive, Honolulu, HI 96822, USA\label{uni:hawaii}
\and 
Institut d’Estudis Espacials de Catalunya (IEEC), E-08034 Barcelona, Spain\label{uni:spain1}
\and
Institute of Space Sciences (ICE, CSIC), Campus UAB, Carrer de Can Magrans, s/n, E-08193 Barcelona, Spain\label{uni:spain2}
\and
Planetary Science Institute, 1700 East Fort Lowell Road, Suite 106, Tucson, AZ 85719-2395, USA\label{uni:baron1}
\and 
Space Telescope Science Institute, 3700 San Martin Drive, Baltimore, MD 21218-2410, USA\label{baltimore}
\and 
INAF - Osservatorio Astronomico di Padova, Vicolo dell’Osservatorio 5, I-35122 Padova, Italy\label{uni:padua}
\and 
Department of Physics, Florida State University, 77 Chieftain Way, Tallahassee, FL, 32306, USA\label{uni:FSU}
\and 
INAF - Astrophysics and Space Science Observatory, Via P. Gobetti 101, 40129 Bologna, Italy\label{uni:pisa}
\and 
School of Physics and Astronomy, University of Leicester, University Road, Leicester LE1 7RH, UK\label{uni:UK}
}

  \date{Received date \today /
   Accepted date } 

\abstract{We present infant-phase observations of the broad-line Type~Ic supernova (SN~Ic-BL)~2020lao, including optical spectroscopy beginning within about 48 hours of the inferred explosion epoch and extending to nearly 100 days. The explosion time was constrained by power-law fits to the rising TESS and ZTF light curves, with the first ZTF detection occurring only $\sim27$ hours after explosion. The optical light curves show a rapid rise that lasted for $\approx \TimeLmax$~days and a peak luminosity typical of SNe~Ic-BL (i.e., $M_r\simeq-18.5$~mag). 
Unlike some engine-driven SN~Ic-BL events, the early light curve of SN~2020lao shows no evidence of an optical afterglow or excess emission, and the absence of any detectable shock–cooling component in the TESS and ZTF data constrains the progenitor to a compact Wolf-Rayet-like star whose $R_\star$ is less than or equal to a few times the $R_\odot$, ruling out any extended envelope. The spectra resemble those of the X-ray-flash-associated SN~2006aj but with systematically higher expansion velocities. From Arnett-type fits to the bolometric light curve and measured \ion{Fe}{ii} $\lambda$5169 line velocities, we infer a $^{56}$Ni mass of $\Nimass~M_{\odot}$, an ejecta mass ($M_{ej}$) of $\Mej M_{\odot}$, and a kinetic energy ($E_K$) of $\sim \EK \times 10^{51}$ erg, corresponding to a specific kinetic energy ($E_K/M_{ej}$) of $\approx \EKoverM\times10^{51}$~erg\,$M_{\odot}^{-1}$. Spectral synthesis modeling broadly reproduces the photospheric-phase spectra of SN~2020lao and suggests $E_K/M_{ej} \approx \EoverMsyn \times 10^{51}$ erg $M_{\odot}^{-1}$.
SN~2020lao and SN~2006aj synthesized comparable amounts of $^{56}$Ni, yet SN~2020lao exhibits $E_K/M_{ej}$ values on the order of 5-10 times larger. Published VLA and \textit{Swift}/XRT non-detections reveal no afterglow emission, allowing us to place stringent limits on relativistic ejecta and dense circumstellar material. Given that SN~2020lao reaches a specific kinetic energy typical of engine-driven SNe~Ic-BL, the lack of an early optical excess together with the non-detections in the radio and X-ray bands suggests that if a relativistic jet was launched, the explosion must have been viewed far off axis or the jet was choked before breakout. If there was no relativistic jet, SN~2020lao would therefore be an extreme nonrelativistic SN~Ic-BL. This underscores the importance of continued infant-phase, multiwavelength monitoring of these explosions.}

\keywords{supernovae : general -- supernovae: individual: SN~2020lao}

\maketitle

\section{Introduction}
\label{sec:introduction}

Historically, the earliest discoveries of exploding stars depended on the detection of high-energy emissions associated with long-duration gamma-ray bursts (GRBs) and X-ray flashes \citep[XRFs;][]{Galama1998,Campana2006,Pian2006,Soderberg2006,Ashall19}. Follow-up observations of such objects can capture an optical and/or radio afterglow accompanied by the emergence of broad-line Type~Ic supernovae (hereafter SNe~Ic-BL; \citealt{Kulkarni1998,Patat2001,Soderberg2006}). Some of these events are thought to be powered by a central engine, originally proposed in the collapsar framework \citep{Woosley1993,MacFadyen1999,Woosley2002a} and more recently in magnetar-driven models \citep{Mazzali2013,Metzger2015,Suzuki2018,Piro2019}, where the collapse of a massive star leaves behind a compact remnant that drives relativistic jets and injects energy into the ejecta. In recent years, an increasing number of SNe~Ic-BL have been discovered through optical transient surveys rather than high-energy emissions \citep[e.g.,][]{Anand2024}. It remains uncertain whether these events produced high-energy emission that went undetected because of observational limitations or unfavorable viewing angles \citep[e.g.,][]{MacFadyen1999,Tanaka2007,Mazzali2005} or arose from a distinct mechanism such as a ``choked'' jet in which the central engine fails to drive a relativistic outflow through the progenitor star \citep[e.g.,][]{Irwin2019,MacFadyen2001,RamirezRuiz2002,Lazzati2012,Piran2019}, or if they instead reflect a fundamentally nonrelativistic explosion.

The advent of robotic telescopes with high-cadence, all-sky imaging capabilities, combined with machine learning techniques for near-real-time transient identification, has ushered in an  era where supernova (SN) discovery rates have soared to thousands per year, with an increasing number detected within hours to a day after the  time of the explosion (hereafter, $t_{exp}$).  For example, the Automated Learning for the Rapid Classification of Events (ALeRCE) identified ZTF20abbplei (hereafter SN~2020lao)  within public stream \textit{Zwicky} Transient Facility (ZTF;  \citealt{Bellm2019}) images taken 2020~May~25.41 UT (i.e., JD~2458994.91; \citealt{Forster2020}). 
As there were already a handful of previous ZTF non-detections, the source was flagged  as a rising SN candidate. A week after discovery, \citet{Burke2020} reported the candidate to be a bona fide SN~Ic-BL. 

SN~2020lao was fortuitously located within the Transiting Exoplanet Survey Satellite (TESS) footprint, with observations covering its position prior to discovery. The full TESS light curve was presented by \citet{Vallely2021}, who inferred a  $t_{exp}$ of 2020 May 25.13 UT (i.e., JD~$2458994.63\pm0.06$). More recently, \citet{Anand2024} estimated a $t_{exp}$  of JD~$2458993.57\pm0.75$ based on power-law fits to the early ZTF $g$- and $r$-band trytry. A reexamination of these datasets (see below) yields a $t_{exp}$ value a day earlier, that is,  on JD~$2458993.64^{+0.23}_{-0.20}$, which is adopted throughout this work.

In this study we analyzed optical broadband photometry and spectroscopy of SN~2020lao beginning within 48 hours of $t_{exp}$. This constitutes a rare dataset of an infant SN~Ic-BL that showed no evidence of high-energy or radio emission \citep{Corsi2023, Anand2024} and no associated optical afterglow component \citep{Burke2020}. 
To verify these absences, we searched archival high-energy transient catalogs over a broad time window around the inferred explosion epoch and found no reported GRB or X-ray transient spatially and temporally coincident with SN~2020lao.
Despite the lack of such high-energy signatures, its light curve shape, luminosity, and spectra closely resemble those of the XRF-associated SN~2006aj (XRF~060218; \citealt{Mazzali2006,Pian2006,Sollerman2006}), albeit with spectral lines suggesting an even more energetic explosion. These similarities, together with the absence of clear engine-driven signatures, raise the question of whether SN~2020lao is related to a GRB viewed off-axis, the outcome of a choked jet, or instead a purely nonrelativistic explosion.
 
\begin{figure}[!ht]
    \centering
    \includegraphics[width=\linewidth]{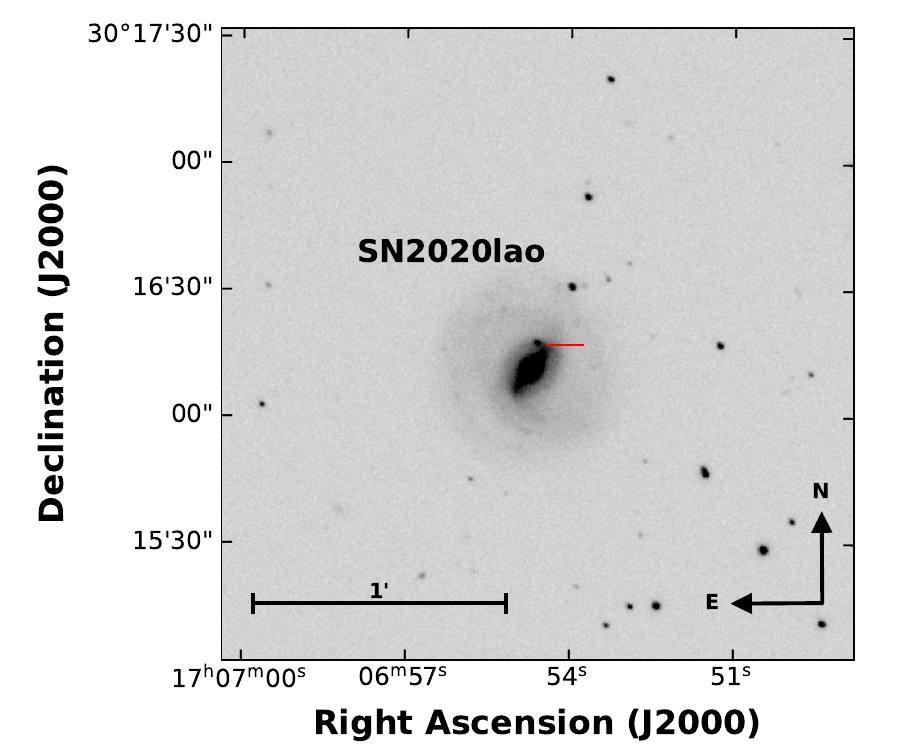}
        \caption{An $r$-band image from the NOT showing the position of SN~2020lao within its host galaxy, CGCG~169-041.}
    \label{fig:FC}
\end{figure}

\begin{figure*}[t]
  \centering
  \sidecaption
  \includegraphics[width=12.0cm]{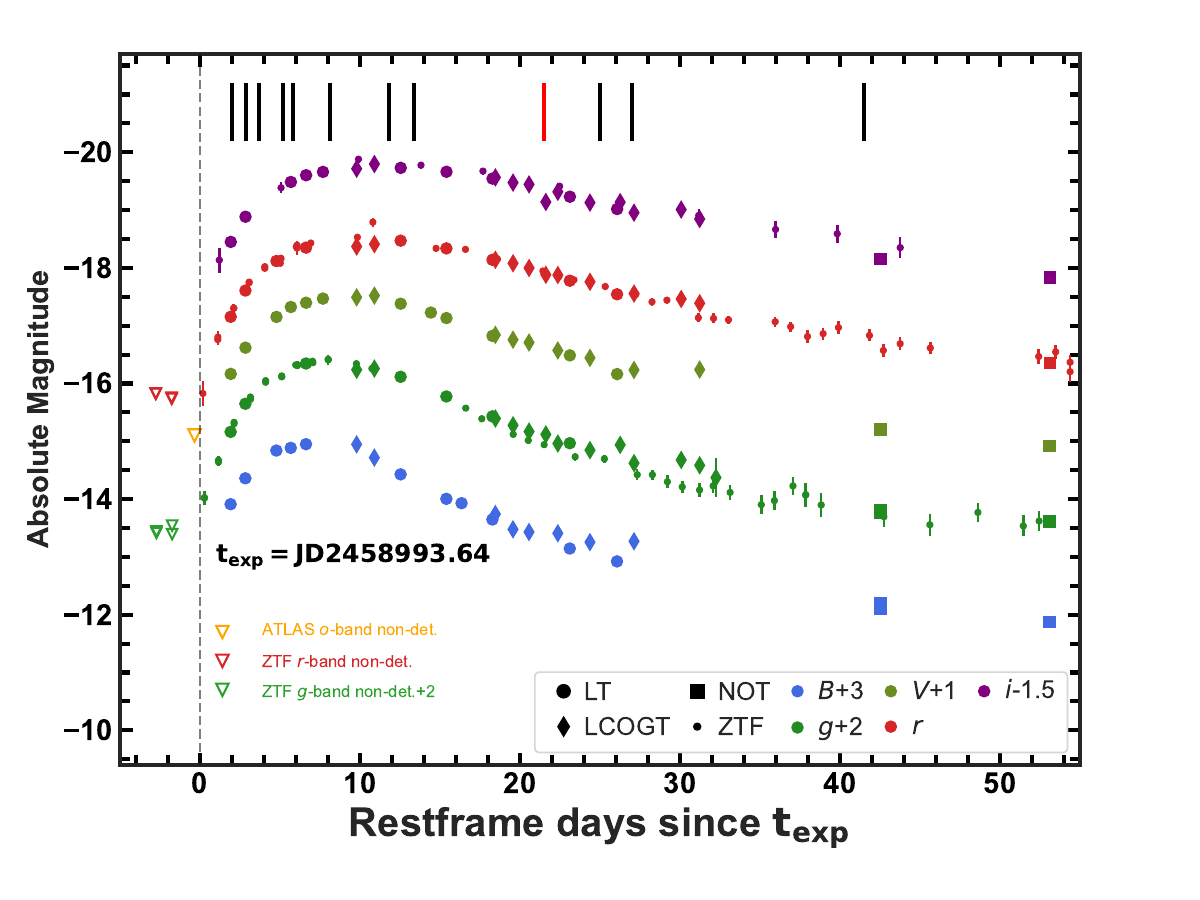}
    \caption{Optical $BgVri$-band light curves of SN~2020lao, corrected for reddening  and vertically offset by arbitrary constants for clarity.  
    The previous ATLAS $o$-band non-detection limit (orange triangle), obtained $-0.34$~days prior to $t_{exp}$ (vertical dashed line), is plotted alongside the $r$-band light curve. ZTF non-detection limits are included. Segments at the top mark the phases when  optical (black) and NIR (red) spectroscopy were obtained (see Table~\ref{tab:specjor}).}
    \label{fig:LCs}
\end{figure*}


\section{Distance, reddening, and host properties} 
\label{sect:distanceANDreddening}

With  J2000 coordinates RA $= 17^{h}06^{m}54^{s}.602$ and Dec $= +30^\circ16^{m}17\farcs35$, SN~2020lao occurred within the face-on, barred spiral galaxy CGCG~169-041 (see  Fig.~\ref{fig:FC}). 
The NASA/IPAC Extragalactic Database (NED) reports a heliocentric redshift of $z = 0.031162 \pm 0.000117$ \citep[e.g.,][]{Marzke1996}. Correcting for the Virgo Cluster, Great Attractor, and Shapley supercluster peculiar velocity model \citep{Mould2000} yields a redshift of $z=0.03245$.
Using cosmological parameters  $\Omega_m = 0.27$, $\Omega_\Lambda = 0.73$ \citep{Komatsu2009}, and $H_0 = 73.30\pm1.04$ km~s$^{-1}$~Mpc$^{-1}$ \citep{Riess2022}, this adjusted redshift corresponds to the adopted luminosity distance of 
$132.7\pm9.3$ Mpc, or  $\mu = 35.61\pm0.15$ mag. 

The Milky Way (MW) reddening in the direction of SN~2020lao is $E(B-V)_{MW}  = 0.044$ mag \citep{Schlafly2011}.  The host-galaxy reddening appears to be minimal, as indicated by the absence of  \ion{Na}{i}~D  in the optical spectra and the post-maximum optical colors (see Sect.~\ref{sec:bblcs}). 

We obtained host-galaxy properties through the \texttt{Blast} website \citep{Jones2024_Blast}\footnote{\href{https://blast.scimma.org/}{https://blast.scimma.org/}.}. \texttt{Blast} estimates galaxy properties through spectral energy distribution (SED) fitting implemented using the \texttt{Prospector}-$\alpha$ model with a nonparametric star formation history (\citealt{Leja2019, Johnson2021_prospector, Wang2023_prospector}; see also Appendix A in \citealt{Jones2024_Blast} for more information). The fitted data include photometry from the Two Micron All Sky Survey \citep{Skrutskie2006}, the Wide-field Infrared Survey Explorer \citep{Wright2010}, the Sloan Digital Sky Survey \citep[SDSS;][]{Fukugita1996, York2000}, and the Panoramic Survey Telescope and Rapid Response System \citep{Chambers2016} using point-spread-function-matched elliptical apertures.

We find the following properties for CGCG~169-041: 
a stellar mass of $\log_{10}(M_\ast/M_\odot) = 11.00^{+0.10}_{-0.05}$, 
a star formation rate of $\log_{10}\left(\mathrm{SFR}/(\frac{M_\odot}{\mathrm{yr}})\right) = 0.09^{+0.20}_{-0.27}$, 
a specific star formation rate of $\log_{10}(\mathrm{sSFR}/M_\odot) = -10.90^{+0.21}_{-0.30}$, 
and a mass-weighted mean stellar age $= 9.02^{+3.26}_{-1.13}$~Gyr. These are consistent with CGCG~169-041 being a young, massive, highly star-forming galaxy.

\section{Observations}

Imaging data were collected with the
Nordic Optical Telescope (NOT) by the NOT Unbiased Transient  (NUTS2)\footnote{\href{https://nuts.sn.ie/}{https://nuts.sn.ie/}.} collaboration,  supplemented with imaging obtained with the Las Cumbres Observatory Global Network (LCOGT), and the Liverpool Telescope (LT). 
  Imaging data were reduced following standard procedures. 
  NOT images were processed using the \texttt{ALFOSCGUI} pipeline\footnote{Alfoscgui is a graphical user interface designed for extracting SN spectroscopy and photometry from data obtained with FOSC-like instruments. It was developed by E. Cappellaro. More information about the package is available at \href{http://sngroup.oapd.inaf.it/foscgui.html}{http://sngroup.oapd.inaf.it/foscgui.html}.},  LCOGT images were retrieved from the LCOGT science archive reduced using the  \texttt{BANZI} pipeline, while fully reduced   images were retrieved from the observatory's data products archive.

\begin{figure*}[!t]
\centering 
\includegraphics[width=\linewidth]{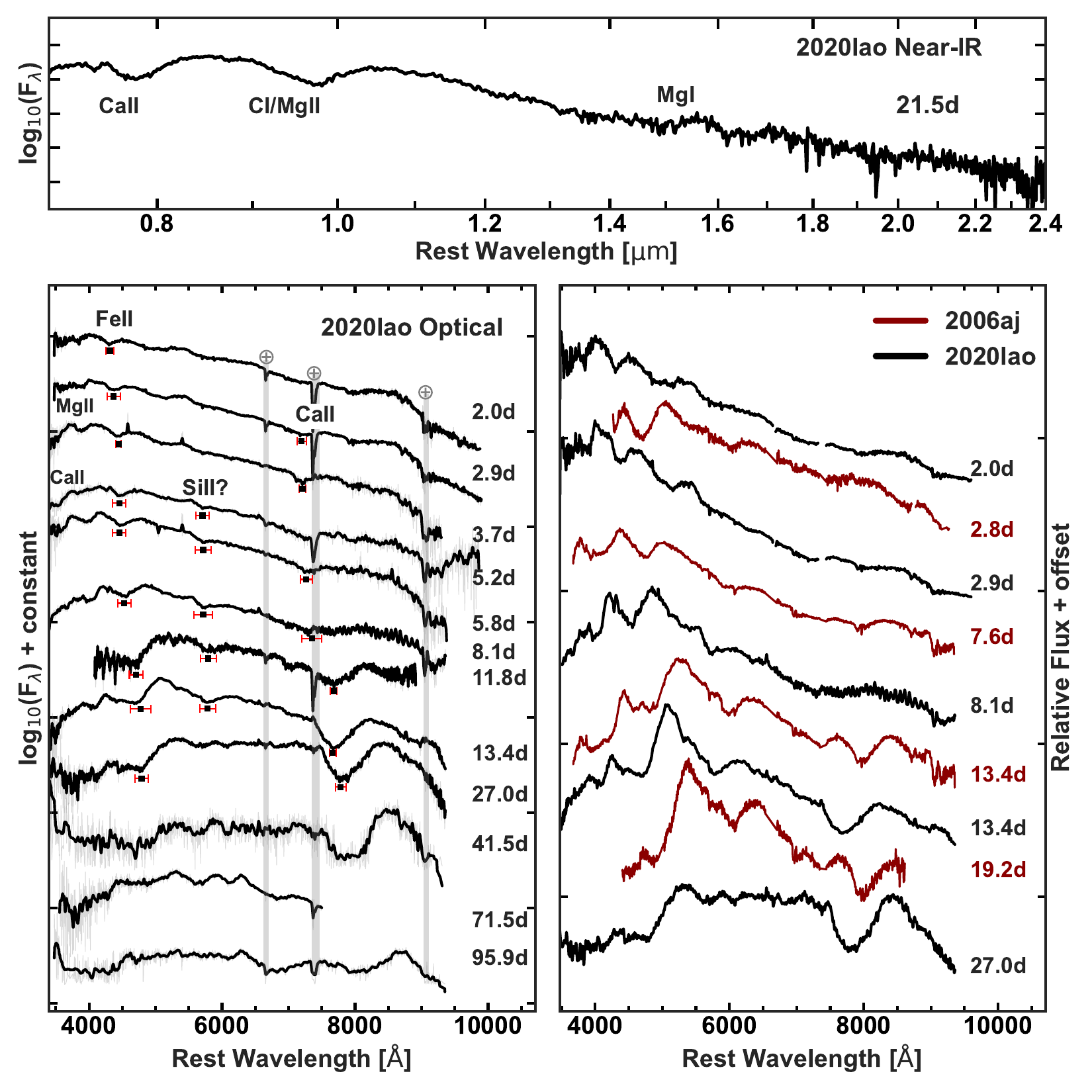}
\caption{Spectroscopic observations of SN~2020lao. \textit{Top:}  NIR $+21.5$~days spectrum of SN~2020lao. \textit{Left:} 
Selected optical spectra of SN~2020lao, corrected for MW reddening and overplotted with a smoothed version. Regions affected by three prevalent telluric features are indicated with vertical gray bands and labeled with Earth symbols. The locations of the absorption minima used to infer the expansion velocities plotted in Fig.~\ref{fig:dopvel} are marked with black dots, with red error bars indicating the wavelength ranges that define the associated uncertainties in the positions of maximum absorption.  \textit{Right:} Comparison between similar epoch spectra of  SN~2020lao and SN~2006aj \citep{Pian2006,Sonbas2008}.}
\label{fig:specseq}
\end{figure*}

Before computing photometry, host-template subtraction was performed on all science images using
\texttt{IMAGEMATCH}.\footnote{\url{https://code.obs.carnegiescience.edu/} } Templates were created by stacking multiple archival LCOGT images from 2014. Point-spread function photometry was then measured using the Aarhus-Barcelona peculiar velocity \href{https://flows.phys.au.dk/}{FLOWS} project's photometry pipeline,\footnote{\url{https://github.com/SNflows}} which employs the effective point-spread-function methodology of \citet{Anderson2000}. Calibration was carried out using nightly zero points, determined by comparing the instrumental photometry of field stars with standard catalog magnitudes from REFCAT2 \citep{Tonry2018b}.

\begin{table*}
\caption{Discovery information.\label{tab:discoveryinfo}}
\begin{tabular}{llll}
\toprule
\toprule
Detection information & JD & $\Delta t_{exp}$\tablefootmark{a} & Magnitude\tablefootmark{b} \\ 
\midrule
\rowcolor{gray!15}
ZTF $r$ non-detection                          & 2458992.81 & $-0.80$ & $> 20.0$\\
ZTF $g$ non-detection                          & 2458992.83 & $-0.78$ & $>20.2$\\ 
\rowcolor{gray!15}
ATLAS $o$ non-detection                        & 2458993.99 & $-0.34$ & $>20.5$\\
TESS Explosion ($t_{exp}$), \citet{Vallely2021}& $2458994.63 \pm 0.06$ & $+0.96$  & $\cdots$\\
\rowcolor{gray!15}
ZTF $t_{exp}$, \citet{Anand2024}                  & $2458993.57 \pm 0.75$ & $-0.07$  & $\cdots$\\
\midrule
Revised TESS $t_{exp}$, this paper  &  $2458993.64^{+0.23}_{-0.20}$ & $~~~0.0$  & $\cdots$\\
\rowcolor{gray!15}
Revised ZTF $t_{exp}$, this paper & 
$2458993.83^{+0.45}_{-0.50}$ & +0.18& $\cdots$  \\ 
ZTF    $r$ recovered detection; this paper    & 2458994.81 & +1.13 & $19.90\pm0.22$ \\ 
\rowcolor{gray!15}
ZTF/ALeRCE $g$ first detection & 2458994.91 & +1.23  & $19.74\pm0.12$\\
\bottomrule
\end{tabular}
\tablefoot{\\
\tablefoottext{a}{Rest-frame days relative to our revised TESS $t_{exp}$ value.}
\tablefoottext{b}{Limiting ZTF magnitudes were computed by the ZTF forced photometry code including the appropriate baseline corrections and adopting a minimum signal-to-noise ratio of 3. ATLAS limiting magnitude as computed by The ATLAS project's forced photometry service.}
}
\end{table*}

Our focused observations are combined with  observations obtained by ZTF.
ZTF $g$- and $r$-band photometry was obtained using the ZTF forced photometry service, with careful implementation of a baseline correction (see \citealt{Shingles2021} for details). Overall, particularly during the earliest phases and through maximum light, the ZTF photometry shows good agreement with measurements computed from our own resources. 
The broadband light curves of SN~2020lao are shown in Fig.~\ref{fig:LCs}.

During our campaign, we obtained thirteen epochs of optical spectroscopy covering a period from +2.0 to +95.9 days after $t_{exp}$ (see Table~\ref{tab:specjor}), including five spectra within the first six days, the earliest of which was published by \citet{Galbany2025}.
 Our first and last two spectra were obtained with the 10.4-m Gran Telescopio Canarias (GTC + OSIRIS), while the others were taken with the 2.56-m NOT (+ ALFOSC), the 2-m LT (+ SPRAT),  the LCOGT 2-m Faulkes telescope (+ FLOYDS), and the 3.5-m Apache Point Telescope (APO) equipped with the Dual Imaging Spectrograph (DIS). The data were reduced following standard procedures \citep[see][]{Hamuy2006}, and the  extracted 1D spectra were color-corrected to match the observed broadband colors of SN~2020lao.
A single near-infrared (NIR) spectrum  was also obtained  at $+$21.5~days with the NASA Infrared Telescope Facility (IRTF; +SpeX, \citealt{Rayner2003}). These data were reduced, calibrated, and telluric corrected following standard SpeXtool procedures with  the telluric corrections accomplished using a spectrum of an A0V star \cite[][]{Cushing2004,Hsiao2019}.  
The spectroscopic sequence of SN~2020lao is plotted in Fig.~\ref{fig:specseq}.\footnote{These data are available in electronic format on the Weizmann Interactive Supernova Data Repository \citep{Ofer2012}.} 

\section{Analysis}

\begin{figure}[!ht]
\centering
\includegraphics[width=\linewidth]{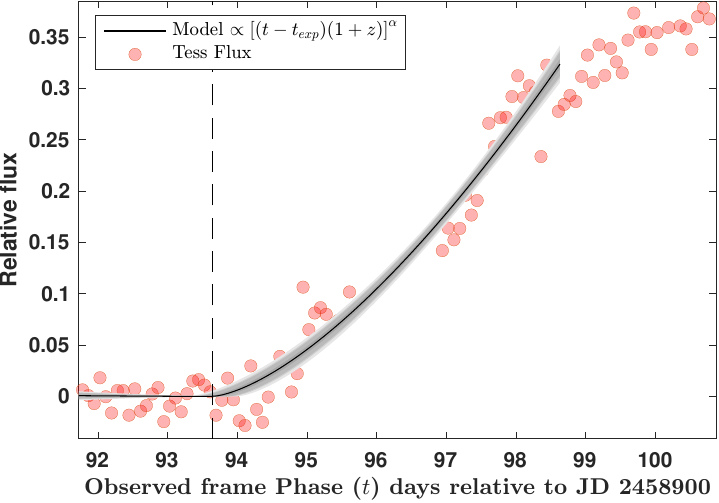}\\
\includegraphics[width=\linewidth]{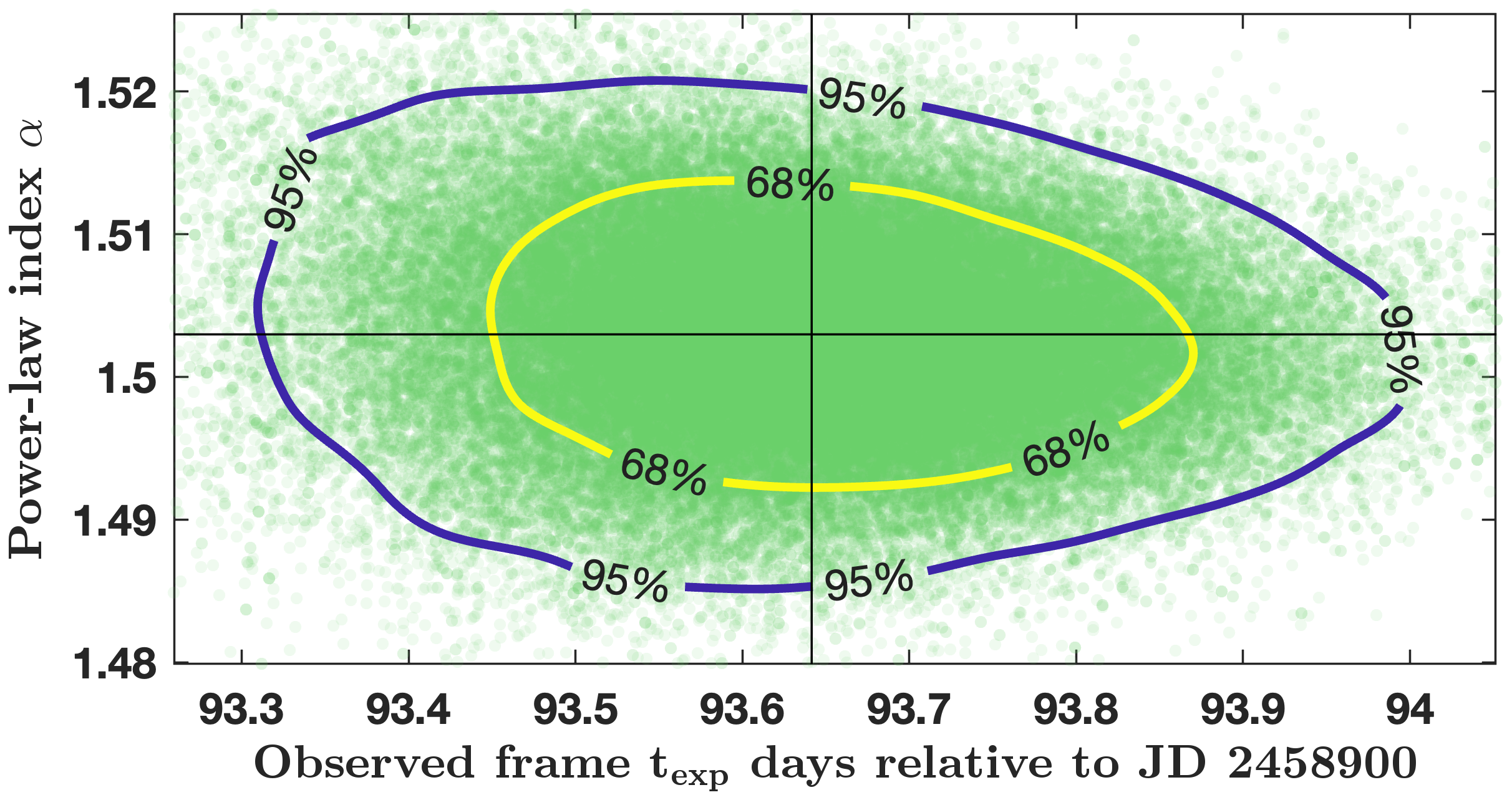}
\caption{\textit{Top panel:} Binned TESS light curve adopting a bin size of two hours and plotted vs. phase relative to JD~2458900. Over-plotted is our best-fit power law. 
 \textit{Bottom panel:} 2D probability density of the MCMC sample between the fit parameters -- time of explosion, $t_{exp}$, and power-law index, $\alpha$. The contours correspond to  68\% and 95\% confidence intervals. The solid intersecting black lines indicate the mean of the MCMC sample for parameters $t_{exp}=\mathrm{JD~}2458993.64^{+0.23}_{-0.20}$ (vertical) and $\alpha=1.50^{+0.01}_{-0.01}$ (horizontal). }
\label{fig:texp}
\end{figure}

\subsection{Early photometry and estimating the explosion epoch}

We used the ZTF forced photometry service \citep{Masci2019} to measure the brightness of SN~2020lao in the ZTF public data stream. The results include a visually confirmed recovered $r$-band detection from the image previously used to compute the original ALeRCE-reported non-detection \citep{Forster2020}. 
The most constraining ZTF non-detections now come from $r$- and $g$-band images obtained 0.80 and 0.78 days before our revised TESS $t_{exp}$, which set limiting magnitudes of $m_r > 20.0$ and $m_g > 20.2$, respectively.  Making use of The ATLAS Project's forced photometry service \citep{Smith2020,Shingles2021}, an even more stringent 5-sigma limiting magnitude of $m_o > 20.5$ mag is measured from a single  $o$-band image taken on  $-$0.34~days (i.e., JD 2458993.99). 
Given the approximately one-day discrepancy between the $t_{exp}$ values computed from the TESS \citep{Vallely2021} and the ZTF \citep{Anand2024} datasets of SN~2020lao, and our recovered $r$-band detection, we revisited the estimate of $t_{exp}$.

To this end, Fig.~\ref{fig:texp} shows the observer-frame TESS light curve of SN~2020lao, along with a single power-law fit to the data spanning the range JD~2458989.13 to JD~2458998.63, after which the light curve shows a prominent break and deviates from a power-law profile. The best-fit model yields a power-law index of $\alpha = 1.50 \pm 0.01$ and an explosion epoch of $t_{exp} =$ JD~$2458993.64^{+0.23}_{-0.20}$. This value precedes the estimate from \citet{Vallely2021}, who inferred $t_{exp} =$ JD~2458994.63 using a different fitting methodology, by 0.96 rest-frame days. Inspection of the TESS light curve (see Fig.~\ref{fig:texp}) shows that at the epoch inferred by \citet{Vallely2021}, the flux is already significantly above the pre-explosion baseline, exceeding the local background by approximately 2.5$\sigma$ and indicating that the SN was already rising at that time.
 We therefore conclude that this later value does not correspond to the true explosion epoch. Our adopted value of $t_{exp} =$ JD\,$2458993.64^{+0.23}_{-0.20}$ is consistent with both the ZTF non-detections and the onset of the rise observed in the TESS light curve.

We also applied the same fitting methodology to the ZTF $r$-band photometry and extended it to JD~2458999.13, which yields $t_{exp} =$ JD~$2458993.83^{+0.45}_{-0.50}$. This value is in agreement with the  JD~$2458993.57\pm0.75$ value reported by \citet{Anand2024}. 

A summary of all non-detections, detections, recovered detections, and $\Delta t_{exp}$ estimates of their corresponding observational dates relative to our revised $t_{exp}$ estimate is provided in Table~\ref{tab:discoveryinfo}. Finally, we note that with our revised TESS $t_{exp}$ value and the recovered ZTF $r$-band detection, the earliest ground-based detection of SN~2020lao occurred $\approx 27$ hours after $t_{exp}$.

\subsection{Broadband light curves}
\label{sec:bblcs}

\begin{figure}[!t]
\centering
\includegraphics[width=0.99\linewidth]{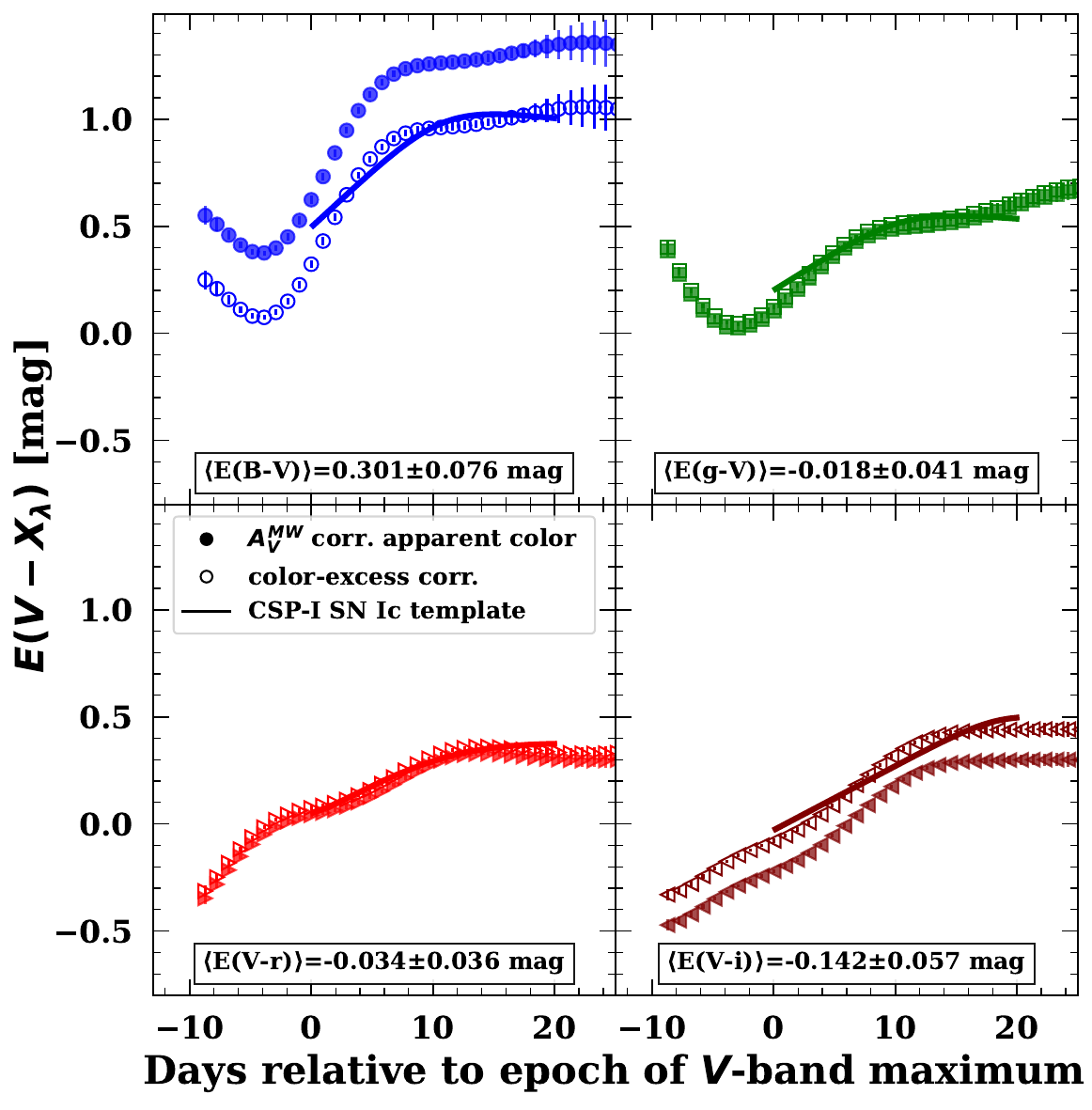}
   \caption{MW reddening-corrected apparent colors of SN~2020lao (filled points) derived from Gaussian process interpolations of the observed light curves. Solid lines represent intrinsic SN~Ic color-curve templates \citep{Stritzinger2018}, while open points show the corrected color curves after adjusting for the average offset between the apparent colors and the templates. The colors of SN~2020lao  indicate that it experienced minimal host-galaxy reddening.} 
    \label{fig:color}
\end{figure}

The light curves of SN~2020lao plotted in Fig.~\ref{fig:LCs} exhibit a bell-shaped profile typical of stripped-envelope (SE) SNe and are characterized by relatively short rise times ($t_{rise}$) to peak. Furthermore, the early light curve evolution shows no evidence of early excess emission, which could be linked to either post-shock breakout cooling or afterglow emission associated with central engine activity. Gaussian process spline fits were used to determine the peak light curve parameters (i.e., time of peak, apparent magnitude at peak, and $t_{rise}$), which are summarized in Table~\ref{tab:peak}. 
The $t_{rise}$ values range between +8.5 days and +11.7 days. In context, they are similar to the rise time of SN~2006aj \citep[e.g.,][]{Sollerman2006}, as well as the average  SDSS-II SN~Ic sample ($t_{rise} \approx 11.5 \pm 0.5$ days). However, the values are somewhat shorter than the averages reported for the SDSS-II SN~Ic-BL  ($t_{rise} \approx 14.7 \pm 0.2$ days; \citealt{taddia2015}) and the intermediate Palomar Transient Factory (iPTF) SN~Ic-BL ($t_{rise} \approx 15 \pm 6$ days; \citealt{taddia2019}) samples. 

\begin{table}
\centering
\tiny
\caption{Peak light curve parameters and rise times.\label{tab:peak}}
\begin{tabular}{clccl} 
\toprule
\toprule
Filter & JD\,2450000 &   $m_{max}$ & $M_{max}$\tablefootmark{a} & $t_{rise}$\tablefootmark{b}   \\
 &  &  (mag) & (mag) & (days)   \\
\midrule 
\rowcolor{gray!15}
$B$ & $9002.42\pm0.09$ &$17.82\pm0.05$ & $-17.97\pm0.35$ &$+8.5\pm0.25$\\
$g$ & $9002.79\pm0.08$ &$17.36\pm0.05$ & $-18.40\pm0.35$ &$+8.9\pm0.24$\\
\rowcolor{gray!15}
$V$ & $9004.20\pm0.10$ &$17.21\pm0.05$ & $-18.54\pm0.35$ &$+10.2\pm0.25$\\
$r$ & $9004.47\pm0.19$ &$17.18\pm0.05$ & $-18.53\pm0.34$ &$+10.5\pm0.21$\\
\rowcolor{gray!15}
$i$ & $9005.69\pm0.19$ &$17.44\pm0.05$ & $-18.24\pm0.34$ &$+11.7\pm0.21$\\
\bottomrule
\end{tabular}
\tablefoot{\\
\tablefoottext{a}{The quoted uncertainties were computed by adding in quadrature errors due to the apparent magnitude estimate, a 0.15 mag systematic uncertainty in the distance modulus, a 0.02 mag systematic uncertainty in $E(B-V)$, and an additional 0.3 mag to account for systematics related to K-corrections \citep[see][]{taddia2019}.}
\tablefoottext{b}{Rest-frame days since $t_{exp}$.}
}
\end{table}

Figure~\ref{fig:color} presents the interpolated optical broadband colors of SN~2020lao, corrected for Galactic extinction, and plotted alongside the intrinsic color-curve SN~Ic templates from \citet{Stritzinger2018}. The colors initially transition rapidly from red to blue as the SN nears peak brightness. This is followed by a gradual shift back to red colors as the ejecta expand, cool, and the photosphere recedes into progressively deeper layers. The color excess values shown in each panel of Fig.~\ref{fig:color} correspond to the average difference between the MW-corrected observed colors and the intrinsic template curves. Fitting these four color excess values simultaneously with the \citet{Fitzpatrick1999} reddening law, and assuming a host-galaxy total-to-selective reddening parameter of  $R^{host}_V = 3.1$, suggests that SN~2020lao suffers minimal host reddening.

Accounting for Galactic reddening and assuming $\mu =35.6$ mag, the peak $r$-band apparent magnitude of SN~2020lao corresponds to $M_r = -18.52\pm0.34$ mag (see Table~\ref{tab:peak}). This value is in agreement with the average value $M_r \sim -18.6\pm0.5$ mag computed from the iPTF sample of three dozen SNe~Ic-BL \citep{taddia2019}.  

\subsection{Spectroscopy}
\label{sec:spectroscopy}

The subset of our spectroscopic time series presented in the left panel of Fig.~\ref{fig:specseq} spans from +2 to +97 days relative to $t_{exp}$. The time series consists of an early high-cadence sequence, with six spectra obtained within the first week. The spectra are characterized by a relatively blue continuum, with the first spectrum well fit by a 
Planck function with a blackbody temperature of 8{,}000~K.
Superposed on the continua of the earliest spectra are several broad absorption troughs, initially centered near 3900~\AA, 4200~\AA, 4500~\AA, 5500~\AA, and 7500~\AA. These features, labeled in Fig.~\ref{fig:specseq}, are attributed to \ion{Ca}{ii} H\&K $\lambda3945$, \ion{Mg}{ii} $\lambda4481$, \ion{Fe}{ii} multiplet 42, and the \ion{Ca}{ii} $\lambda\lambda8498, 8542, 8662$ NIR triplet, while the trough near 5500~\AA, first seen in the $+5.3$~day spectrum, is tentatively associated with \ion{Si}{ii} $\lambda6355$.
As the spectra evolve, the pseudo–equivalent width of the \ion{Ca}{ii} NIR triplet increases markedly between $+8.1$ and $+13.4$~days and remains prominent at later epochs.

\begin{figure}[!ht]
    \centering
    \includegraphics[width=\linewidth]{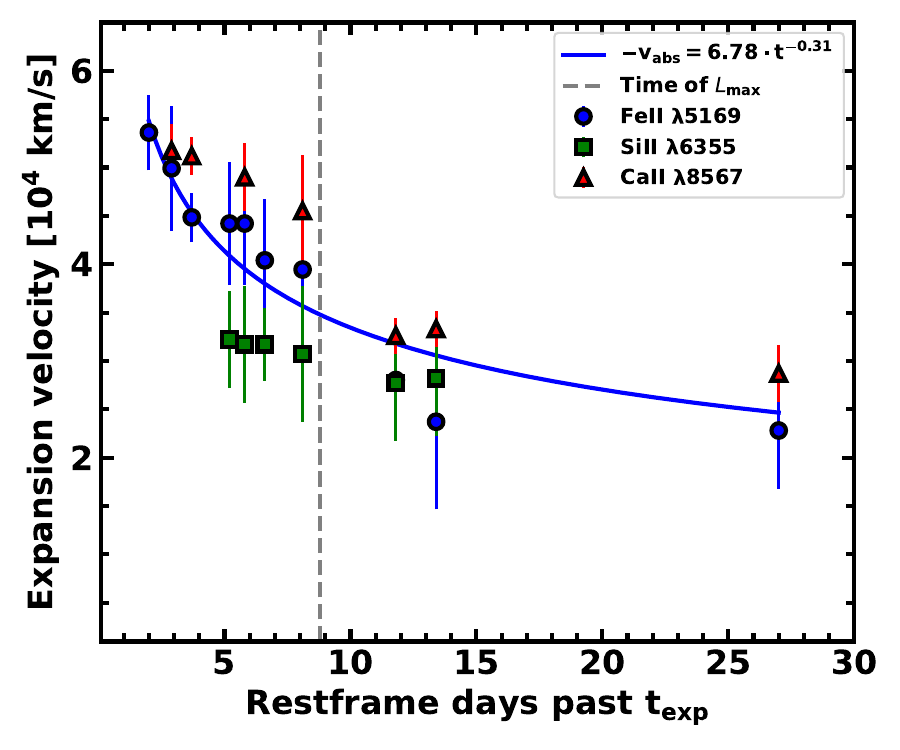}
    \caption{Doppler velocity evolution as measured from the position of maximum absorption for \ion{Fe} {ii}~$\lambda5169$, \ion{Si}{ii} $\lambda6355$, and  \ion{Ca}{ii} $\lambda$8567.}
    \label{fig:dopvel}
\end{figure}

Attributing the 4300–4500~\AA\ absorption feature in the earliest spectra of SN~2020lao to $\ion{Fe}{ii} \lambda$5169, the expansion velocity from the position of maximum absorption ($-v_{abs}$) is inferred using the relativistic Doppler formula. From the minimum of the feature in the $+$2~days spectrum, this yields $-v_{abs}(\ion{Fe}{ii}\ \lambda$5169) $\approx 53{,}600 \pm 3{,}900$ (random) $\pm 8{,}000$ (systematic) \kms. The velocity decreases to $\approx  22{,}800 \pm 4{,}950\, \mathrm{(random)} \pm 8{,}000 \, \mathrm{(systematc)}$ \kms\ by +27~days. The adopted value of 8{,}000~km~s$^{-1}$ for the systematic uncertainty reflects the effects of line blending, trough identification, and uncertainties associated with the choice of reference wavelength for the \ion{Fe}{ii} multiplet~42. Although the absorption feature near 5000~\AA\ may include contributions from the $\lambda\lambda$4924, 5018, and 5169 components \citep[see][]{Branch2002}, it is more commonplace in the literature to adopt $\lambda$5169 as the rest wavelength when measuring photospheric velocities in SNe~Ic-BL (e.g., \citealt{Modjaz2016,taddia2019,Williamson2023,Finneran2025}).
If instead $\lambda$5018 were adopted as the reference wavelength, the inferred values of $-v_{abs}$(\ion{Fe}{ii}) would be systematically lower by $\approx$ 8{,}000~\kms, comparable to the magnitude of the adopted systematic term \citep[see][]{Prentice2017}.
Fitting a power-law function to the $-v_{abs}$(\ion{Fe}{ii} $\lambda5169$) evolution and evaluating the fit at the time of  
  peak luminosity  ($L_{max}$), we infer
$-v_{abs}(\ion{Fe}{ii}\ \lambda5169) \approx \FeIIVelPeak~\kms$.
This value is adopted below  to estimate the specific energy of SN~2020lao.  

Figure~\ref{fig:dopvel} shows the evolution of 
$-v_{\mathrm{abs}}$(\ion{Fe}{ii}), together with the corresponding 
$-v_{\mathrm{abs}}$(\ion{Si}{ii}) and $-v_{\mathrm{abs}}$(\ion{Ca}{ii}), as 
measured from the minima of the absorption features highlighted in the left  panel of Fig.~\ref{fig:specseq}.
If correctly identified as \ion{Si}{ii}, the feature implies pre-maximum velocities lower than those from \ion{Fe}{ii}; later, more reliable identifications yield velocities comparable to both the \ion{Fe}{ii} and the \ion{Ca}{ii} NIR triplet features.

The single NIR spectrum of SN\,2020lao, spanning 7{,}000–25{,}000~\AA\ and shown at the top of Fig.~\ref{fig:specseq}, is dominated at shorter wavelengths by two broad absorption features. These include the \ion{Ca}{ii} NIR triplet, also present in the optical spectra, and a prominent feature with an absorption minimum near $\approx 1.0\, \mu$m, likely formed by \ion{C}{i} $\lambda 1.0693\, \mu$m with possible contributions from \ion{He}{i} $\lambda 1.0830 \, \mu$m and \ion{O}{i}\, $\lambda 1.1290\, \mu$m. The Doppler velocity derived from the \ion{Ca}{ii} minimum is consistent with the velocity evolution measured from the optical spectra. A third prominent feature in the spectrum, peaking just blueward of $1.6, \mu$m, is likely formed by the blending of \ion{Mg}{i} $\lambda 1.4878\, \mu$m and \ion{Mg}{i} $\lambda 1.5033\, \mu$m. Beyond the main spectral features, numerous weaker absorption lines can be identified across the spectrum, in agreement with the line identifications reported for SN~2013ak by \citet{Shahbandeh2022}. Finally, the absence of the $\ion{He}{i}\, \lambda 2.0581\, \mu$m feature in the NIR spectrum further supports the classification of SN~2020lao as a He-poor SE~SN. 

\section{Discussion}

\subsection{Estimating the explosion parameters of SN~2020lao}

\subsubsection{Bolometric light curve and Arnett-model fits}
\label{sec:explosionparameters}

\begin{figure}
    \centering
     \includegraphics[width=\linewidth]{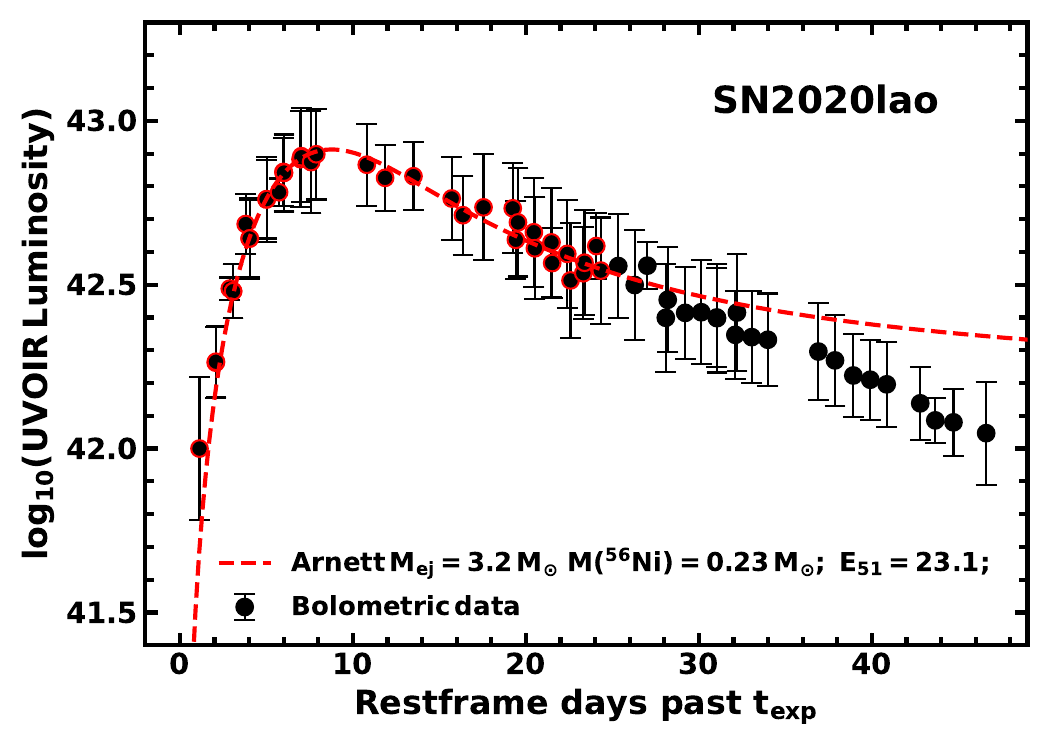}
    \caption{\citet{Arnett1982} model fit to the bolometric light curve of SN~2020lao constructed by integrating blackbody functions matched to the observed SEDs.
    Explosion parameters derived from the fit, with $E_K$ inferred from the $E_K/M_{\rm ej}$ relation of \citet{Wheeler2015}, are provided.}
    \label{fig:arnett}
\end{figure}

Explosion parameters were estimated by fitting the pseudo-bolometric light curve of SN~2020lao with the analytical model of \citet{Arnett1982}. The pseudo-bolometric light curve was constructed using the Python package \texttt{superbol} \citep{Nicholl2018}. 
As a first step, the broadband light curves were fit with polynomial functions, which were used to generate interpolated photometry at epochs with missing coverage in the reference band (chosen to be the $r$ band). The photometry was   corrected for MW reddening  and then converted to flux. Next, observed SEDs were constructed for each epoch $r$-band photometry was obtained, and a blackbody function was fit to each of the observed SEDs. The resulting blackbody fits were integrated over wavelength to obtain the ultraviolet, optical, and infrared (UVOIR) flux, which was converted to luminosity using the adopted distance to the host galaxy. 
The resulting pseudo-bolometric light curve is shown in Fig.~\ref{fig:arnett}, where the associated error bars account integration error. 

To estimate the $^{56}$Ni abundance powering the emission of SN~2020lao and the  ejecta mass ($M_{ej}$) of SN~2020lao a \citet{Arnett1982}-like analytical  model presented by \citet[][see their Eq. 1]{Cano2013}, was fit to the UVOIR light curve,  while the explosion energy ($E_K$) was inferred making use of  relation $E_K / M_{ej} = \frac{3}{10} v_{ph}^2$ \citep{Wheeler2015}.\footnote{The parameter $v_{ph}$ in this relation is taken to be equivalent to $-v_{abs}(\ion{Fe}{ii}\ \lambda5160$) at the time of $L_{max}$.} 
Although this analytical method relies on simplifying assumptions including spherical symmetry, homologous expansion with a fixed density profile \citep{taddia2018}, centrally concentrated $^{56}$Ni, and constant  optical opacity ($\kappa$), it provides a practical baseline for comparison with larger SE SN samples \citep[e.g.,][] {Drout2011,Lyman2016,Prentice2017,taddia2019} and with results from more sophisticated modeling approaches \citep[e.g.,][]{taddia2018}.

The Arnett model fit to the UVOIR bolometric light curve over the first
$\sim 25$ days is shown in Fig.~\ref{fig:arnett}, adopting a constant opacity of
$\kappa = 0.07~{\rm cm^2\,g^{-1}}$ \citep{Chugai2000}. The best-fit model yields a
rise time of $t_{\rm rise} = \TimeLmax$~days and a peak luminosity of
$\log_{10}(L_{\max}) = 42.913$, corresponding to a synthesized $^{56}$Ni mass of
$\Nimass\,M_{\odot}$ after accounting for uncertainties in both the light-curve
fit and the adopted distance. The inferred ejecta mass is
$M_{\rm ej} = \Mej\,M_{\odot}$, with an uncertainty reflecting contributions from
the peak photospheric velocity and the fitting procedure.

As detailed in Sect.~\ref{sec:spectroscopy}, the photospheric velocity at peak light,
$v_{\rm ph} \simeq \FeIIVelPeak~{\rm km\,s^{-1}}$, is obtained from a power-law
fit inferred from the position of  the at maximum absorption and assuming a reference wavelength  of \ion{Fe}{ii}  $\lambda5169$ 
(Fig.~\ref{fig:dopvel}). Using this velocity together with the specific kinetic
energy relation from \citet[][their Eq.~3, 
$E_{K}/M_{\rm ej} = \tfrac{3}{10}v_{\rm ph}^2$]{Wheeler2015}, we infer  $E_K/M_{ ej} = \EKoverM\times 10^{51}$~erg~$M_\odot^{-1}$.
Substituting the Arnett-model fit inferred $M_{ej}$ yields an 
$E_K \simeq \EK \times 10^{51}$~erg, where the quoted uncertainty includes the errors in both $v_{\rm ph}$ and $M_{\rm ej}$.

We emphasize that the $E_K$ and $M_{ej}$ parameters depend on the
assumption of a constant opacity, $\kappa$. As demonstrated by \citet{taddia2018}, this
approximation is accompanied with  systematic uncertainty for
SE SNe: varying $\kappa$ across a plausible range 
($0.05$--$0.15~{\rm cm^2\,g^{-1}}$) can change the inferred $M_{ej}$ and
$E_K$ by factors of $\sim 2$. 
The reported values of $M_{\rm ej}$ and $E_K$ based on Arnett model fitting  should be interpreted with this inherent,
model-dependent uncertainty in mind.

\subsubsection{SN~2020lao in the context of the SN~Ic-BL paradigm and spectral synthesis}
\label{sec:spectralsynthesis}

As a starting point, we must set aside the overly simplistic view that all  SNe~Ic-BL are spectroscopically equivalent. In practice, the amount of line blending spans a wide range from a  few thousand to several tens of thousands of \kms, yielding strikingly different spectral characteristics, as demonstrated by \citet{Prentice2016}.
 These authors suggested a very simple classification scheme for SNe\,Ic, based simply on the number of spectral features observed near maximum light in the range 4000 -- 7500\,\AA. The number of features shown by SNe\,Ic in this portion of the optical spectrum ranges from n=7 in what we could call ``narrow-line'' objects to as few as n=3 for the very  SNe~Ic BL  associated with GRBs. However, it is customary to call all SNe~Ic with $n<7$ ``broad-lined,'' which is  not very precise. A smaller value of $n$ typically indicates a larger   $E_K/M_{ej}$ ratio, although even this assumption does not fully describe the range of possibilities. In fact, the value of $n$ crucially depends on the density slope in the line-forming layers. 
For example, SN\,1998bw has $n=3$, and so do all well-observed GRB-associated SNe~Ic-BL, but the XRF SN\,2006aj has $n=6$, and blending only affects the \ion{Fe}{ii} multiplet 42 lines. Fewer features means that the line-forming region spreads over a larger range of velocities, such that some material reaches high enough velocities that lines formed there can appear blueshifted and blend with the next strong line located at adjacent blue wavelengths. As the separation in velocity between the various features is not constant, ejecta reaching progressively larger velocities show a higher degree of feature blending. It takes ejecta of $v_{ph} \sim 30{,}000$\,\kms\ to blend the \ion{Ca}{ii} NIR triplet with \ion{O}{i} $\lambda$7774, feature such that only three major features are observed  in known GRB-associated SNe~Ic-BL. 

\begin{figure}
\centering
\includegraphics[width=0.48\textwidth]{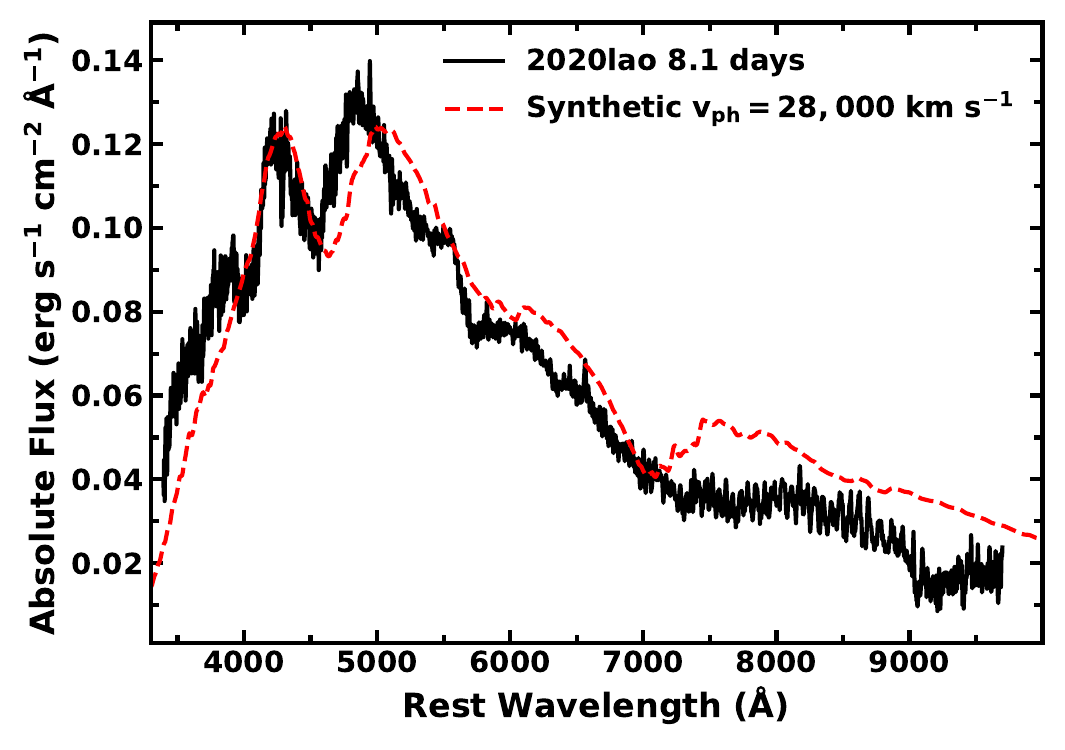} 
\includegraphics[width=0.48\textwidth]{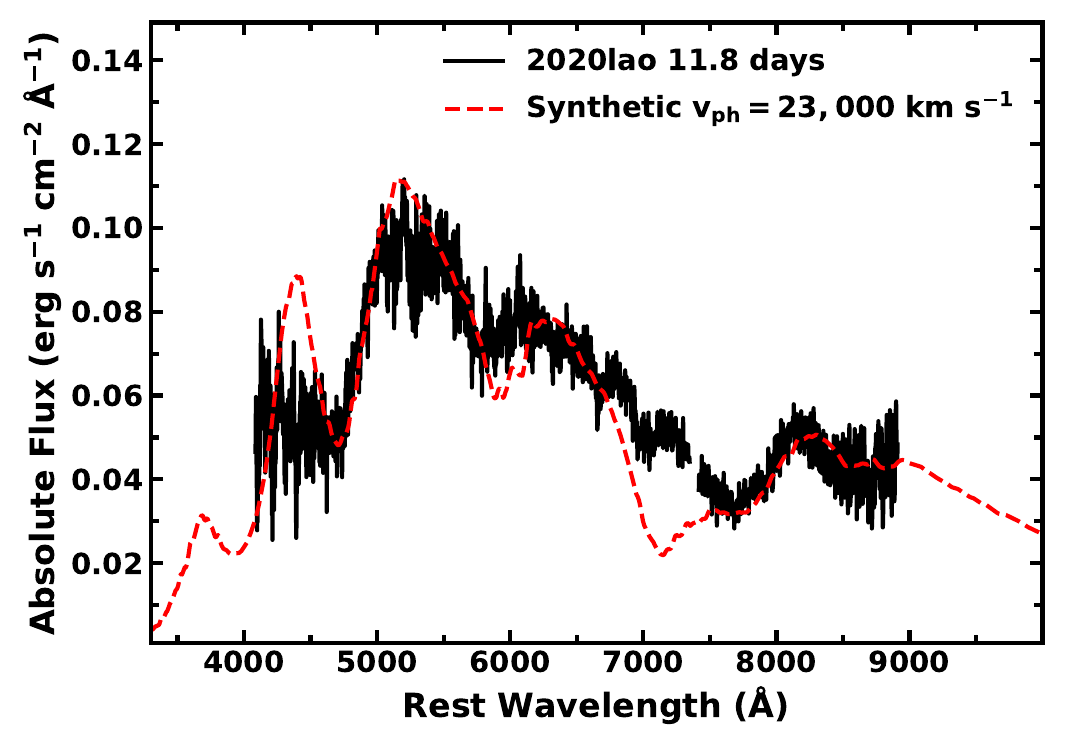} 
\includegraphics[width=0.48\textwidth]{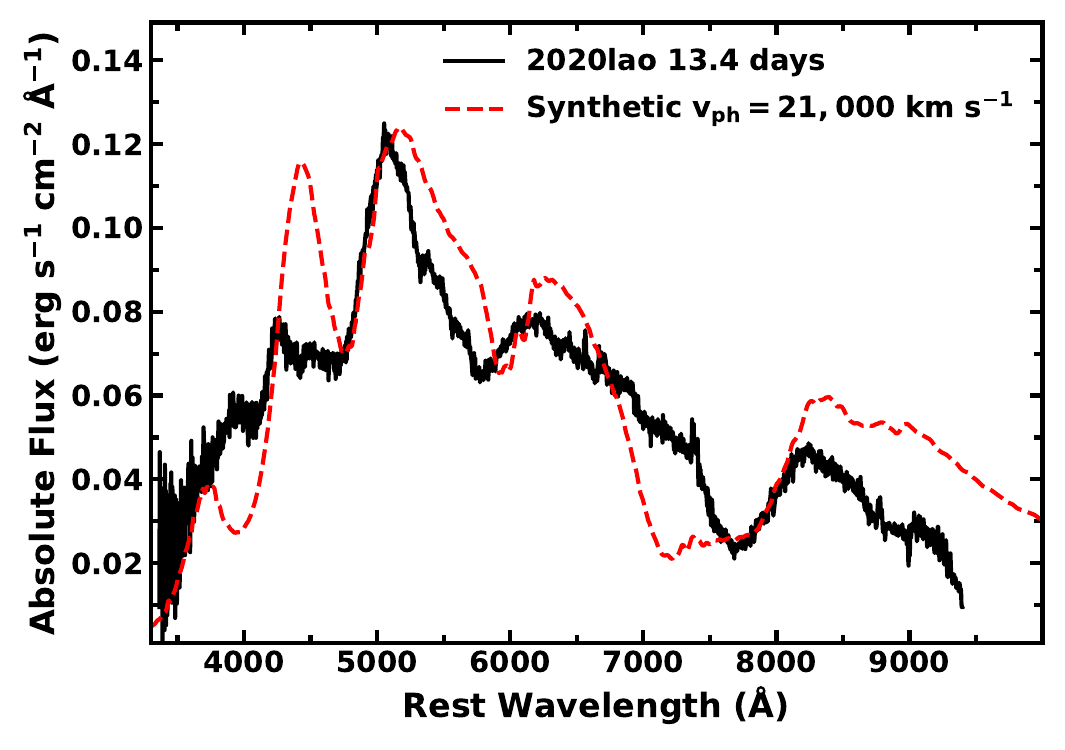} 
\caption{Synthetic spectra from a model with $M_{ej} = \Mejsyn M_{\odot}$ and $E_{K} = \EKsyn \times 10^{51}$ erg compared to observed spectra of SN~2020lao at $+8.1$ (top), $+$11.8 (middle), and  $+13.4$~days (bottom) past explosion. }
\label{fig:synspectra}
\end{figure}

Furthermore, \citet{Mazzali2017} discussed at length how the structure of
the outer layers influences the spectral appearance. In particular, a
different density slope in the outer layers has a significant impact on the appearance
of the early spectra, including the number of lines. A flatter density
distribution $(\rho(v))$ leads to broader lines, and therefore a lower value of
$n$. The value of $n$ can also change with time as the line-forming region
recedes to inner layers of the ejecta. Alternatively, an increase in mass that is \emph{not} accompanied by a flattening of $\rho(v)$ produces a ``fast-lined'' spectrum, in which the intrinsic line widths remain unchanged but the blueshifts of the features become larger.

This is apparently the case for SN\,2020lao, which  could therefore be called a ``fast-lined'' SN~Ic-6, rather than just, generically, a SN~Ic-BL. A similar case was that of PTF12gzk \citep{Ben-Ami2012}, when compared, for example, to SN\,2004aw \citep{Taubenberger2006}. 

These considerations are important when one tries to ``rescale'' an explosion model (i.e., a density structure) used for a particular SN based on its light curve and spectra to match the properties of another SN. As explained in \citet{Mazzali2017}, this should be attempted only for SNe that look rather similar. In the particular case of SN\,2020lao, the light curve evolves similarly to that of SN\,2006aj, also as we discuss below, achieves a similar peak luminosity, while the spectra are similar in shape to those of SN\,2006aj but with features exhibiting much higher velocities. This clearly calls not for a simultaneous increase in both $M_{ej}$ and $E_k$ via a flattening of the outer layers, which is the trend that characterizes SNe-Ic with progressively smaller $n$ values and translates into two separate contributions to $E_K$, but only for a shifting of the line-forming region to higher velocities, starting from the model used for SN\,2006aj. The steeply declining density of the outer ejecta that characterize the model should be preserved in order not to change the blending of the spectral features. Actually, this is basically the assumption of the popular Arnett approach \citep{Arnett1982}.  

Although the spectra of SN\,2020lao are morphologically similar to those of SN\,2006aj, the lines in SN\,2020lao are systematically faster. The light curves are also very similar (see below), apart from a modest difference in luminosity. At the time of $L_{\max}$, the position of the \ion{Fe}{ii}\ $\lambda5169$ absorption minimum in SN~2020lao indicates $-v_{\rm abs}(\ion{Fe}{ii}\ \lambda5169) \approx$ \FeIIVelPeakVELONLY\ \kms, while in the case of SN\,2006aj, \citet{Maund2007} reported  a substantially  lower value of  $-v_{\rm abs}(\ion{Fe}{ii}\ \lambda5169) \approx 18{,}500\pm3{,}000$ \kms\ \citep[see also][]{Lyman2016}. 

This comparison suggests that a suitably rescaled version of the explosion model used for SN~2006aj can account for SN~2020lao, provided that the steeply declining outer ejecta are located approximately 10{,}000\,\kms\ farther out. This requirement implies that SN~2020lao has both a larger $E_{K}$ and a larger $M_{\rm ej}$, but not a flatter density ($\rho(v)$) profile in the outer layers. Using a simple Arnett-type rescaling and directly comparing the properties of the two SNe~Ic-BL, and assuming that the opacity is the same in the two objects rather than estimating the actual time- and depth-dependent opacity, we infer approximate values for SN~2020lao of $M_{\rm ej} \approx 3.5\,M_{\odot}$ and $E_{K} \approx \EKsyn \times 10^{51}$ erg.
 
Spectral modeling was performed on three spectra of SN~2020lao, which enabled us  to assess whether the previously estimated properties of the SN are consistent with its observed spectral features.  To this end, a 1D Monte Carlo radiative-transfer  code developed by \citet{mazzali1993}, \citet{lucy1999}, and \citet{Mazzali2000} was  employed.  The code has previously been used to analyze the spectra of both thermonuclear and core-collapse SNe, including among others, the SNe~Ic-BL 2006aj \citep{Mazzali2006} and 2016jca \citep{Ashall19}. The method is particularly well suited for generating synthetic spectra during the photospheric phase, when the inner ejecta are optically thick and a well-defined photosphere exists.
Assuming that the photosphere emits a blackbody spectrum, the spectral synthesis code releases photon packets from the photosphere and computes their subsequent propagation through the SN ejecta. The code requires as input the density structure, abundances, the location of the photosphere, and the bolometric luminosity. 

In this work,   a uniform abundance distribution is assumed with a composition of 82\% oxygen, 16\% neon,  2\% silicon, and 0.2\% carbon.  This composition is consistent with that of the carbon-oxygen  layers in massive stars \citep[e.g.,][]{Woosley2002b}. Additional notably trace elements  in the ejecta include 0.8\% magnesium, and 0.0005\% calcium. Motivated by the spectroscopic similarity between SN~2020lao and SN~2006aj (see Fig.~\ref{fig:specseq}, right), we adopted the $\rho(v)$ used in the spectral modeling of SN~2006aj \citep{Mazzali2006}, scaled appropriately to compute synthetic spectra for SN~2020lao. A discussion of the uncertainties associated with this approach can be found in \citet{Ashall20}. The bolometric luminosity of SN~2020lao was taken from the Arnett-model results presented in Sect.~\ref{sec:explosionparameters}.

Figure~\ref{fig:synspectra} shows the comparison between the observed spectra of SN~2020lao at +8.1, +11.8, and +13.4~days and the corresponding synthetic spectra computed for $E_K/M_{ej} = \EoverMsyn \times 10^{51}$~erg $M_{\odot}^{-1}$ and photospheric velocities of 28{,}000, 23{,}000, and 21{,}000~\kms, respectively, with uncertainties on the order of up to 10\%. Overall, the observed spectra are reasonably well reproduced, with all major absorption complexes present in the models. The most notable discrepancy is the broad feature between 7000 and 8000~\AA\ in the later spectrum where the synthetic spectra show a more pronounced blend of \ion{O}{i}~$\lambda7773$ and the \ion{Ca}{ii} NIR triplet than is seen in the observations. This discrepancy may stem from the assumption of a uniform abundance distribution, or  the fact that the underlying ejecta may be aspherical.

Having established the overall spectral agreement, we turned to the $v_{ph}$ values inferred from the spectral modeling. A least-squares linear fit to the model-derived velocities allowed us to estimate the photospheric velocity at the time of $L_{\max}$. The extrapolated value is $v_{ph} = 27{,}000$~\kms, which is $\sim 7{,}000$~\kms\ lower than the $-v_{abs}(\ion{Fe}{ii}\ \lambda5169)$ estimate, although still consistent within the adopted uncertainties (see Sect.~\ref{sec:spectroscopy}).

With an estimate of $v_{ph}$ in hand, we proceeded to compare the synthetic spectrum inferences of SN~2020lao for $E_K$ and $E_K/M_{ej}$ to those inferred in Sect.~\ref{sec:explosionparameters}, based on a combination of the Arnett-model fit, our estimate of $-v_{abs}(\ion{Fe}{ii}\ \lambda5169)$, and the relation $E_K/M_{ej} = \frac{3}{10}v_{ph}^2$. We find that this method provides $E_K$ and $E_K/M_{ej}$ values that are  $\approx 36\%$ and $47\%$ larger, respectively.

Finally, to place these results in context, we compared our inferred explosion parameters of SN~2020lao to those of SN~2006aj. Arnett model fits to the bolometric light curves of both objects indicate that they synthesized approximately $0.2\ M_{\odot}$ of $^{56}$Ni \citep{Mazzali2006,Cano2013}. Spectral synthesis modeling of SN~2006aj suggests $v_{ph} = 18{,}000$ km s$^{-1}$ (comparable to the value inferred from its maximum light spectra, i.e., $-v_{abs}(\ion{Fe}{ii}\ \lambda5169) \approx 18{,}500$ km s$^{-1}$; \citealt{Maund2007}), $E_K = 2\times10^{51}$ erg, $M_{ej} = 2\ M_{\odot}$, and hence $E_K/M_{ej} = 1\times10^{51}$ erg $M_{\odot}^{-1}$. These parameter values are all lower than those inferred from our spectral model fits of SN~2020lao, as well as the even higher values inferred from the Arnett model fitting and the $E_K/M_{ej}$ relation of \citet{Wheeler2015}. Quantitatively, our inferred range of $E_K/M_{ej}$ values for SN~2020lao is on the order of 5-10 times larger than that of SN~2006aj. 
In conclusion, we emphasize that the results from the spectral synthesis models rest on a number of underlying assumptions, several of which may no longer hold if the ejecta of SN~2020lao are significantly aspherical.

\subsection{Constraint on the progenitor's radius}
\label{sec:radius}

The early TESS and ZTF light curves also allowed us to constrain the progenitor radius. To translate the lack of any early optical excess into a quantitative limit, we adopted the analytic shock–cooling scalings for compact, H/He-poor  envelopes  derived from \citet[][their Eq.~15]{Rabinak2011}, where the luminosity is given  by

\begin{equation}
L(t)
= 9.9\times10^{42}\,
\frac{E_{51}^{0.85}\,R_{*,13}}
{f_\rho^{0.16}\,(M_{ej}/M_\odot)^{0.69}\,\kappa_{0.34}^{0.85}}\,
t_{5}^{-0.35}\ \ {\rm erg\,s^{-1}}.
\end{equation}
Here $E_{51}=E_K/10^{51}\,{\rm erg}$, $R_{*,13}=R_\star/10^{13}\,{\rm cm}$, $f_\rho$ encodes the outer density profile (we adopted $f_\rho=0.1$, appropriate for compact WR-like envelopes; see Appendix A of \citealt{Calzavara2004}), $\kappa_{0.34}\equiv \kappa/0.34\,{\rm cm^2\,g^{-1}}$, and $t_{5}\equiv t/10^{5}\,{\rm s}$.  Solving for $R_\star$ and using our explosion parameters from the spectral synthesis analysis ($E_K=\EKsyn \times10^{51}\,{\rm erg}\Rightarrow E_{51}= 17$; $M_{ej} = \Mejsyn~\,M_\odot$) with the first ZTF point at $t=1.13$ days and $\log_{10}L\simeq42.000$, we obtain for $\kappa=0.07$ an $R_\star \lesssim 0.6\ R_\odot$.
Alternatively, adopting $\kappa = 0.20$ yields $R_\star \lesssim 1.4\, R_\odot$, while $\kappa = 0.34$ gives $R_\star \lesssim 2.1\ R_\odot$. At the second epoch ($t=2.07$ days, $\log_{10}L \simeq 42.262$), the corresponding limits are $R_\star \lesssim  1.3\,R_\odot$ for $\kappa = 0.07$ and $R_\star \lesssim 4.8\,R_\odot$ for $\kappa = 0.34$. 

If we instead   require the shock--cooling contribution to be $\le$30\% of the measured luminosity to avoid producing a visible early bump  \citep{Piro2013,Nakar2014}, then the limits scale down linearly with $L$, tightening to $R_\star \lesssim 0.2 - 0.6\, R_\odot$ at +1.13 days and $R_\star \lesssim 0.4 - 1.4\, R_\odot$ at +2.07 days. Even allowing for uncertainties in $\kappa$ or modest bolometric corrections, the data robustly favor a compact Wolf-Rayet (WR)-like progenitor with $R_\star$ well below a few $R_\odot$, and disfavor any extended envelope that could produce an early optical excess. Finally, adopting the explosion parameters inferred from the bolometric light curve and observed spectra  ($E_{51} = 23.1$ erg, and $M_{ej}=3.2$ $M_{\odot}$; see Sect.~\ref{sec:explosionparameters}) suggests even tighter constraints on $R_\star$.   

\subsection{Constraints from radio and X-ray observations}

The optical data of SN~2020lao show no signatures of central-engine activity. The light curves lack any early excess seen in GRB- or XRF-associated SN~Ic-BL events, for example SN~2006aj \citep{Campana2006,Soderberg2006}, SN~2017iuk, \citep{Izzo2019}, and SN~2020bvc \citep{Izzo2020,Ho2020}, and the spectra presented in Fig.~\ref{fig:specseq} identify it as a high-velocity, energetic SN~Ic-BL without an afterglow component. 

Published Very Large Array observations of SN 2020lao yielded 3$\sigma$ non-detections at +13 days (5.2 GHz; $F_\nu \lesssim 33$ $\mu$Jy) and +141 days (5.5 GHz; $F_\nu \lesssim 21$ $\mu$Jy; \citealt{Corsi2023}). 
These are among the faintest radio limits measured for SNe~Ic-BL \citep{Bietenholz2021} and effectively exclude relativistic ejecta similar to those in SN~1998bw \citep{Kulkarni1998}, unless the circumburst density was unusually low.
For a distance of $D=132.7$~Mpc, the observed limits correspond to 
luminosity limits of $L_\nu \lesssim 6.9\times10^{26}$ and $4.4\times10^{26}$~erg~s$^{-1}$~Hz$^{-1}$ at 5.2 GHz and 5.5 GHz, respectively. 

\begin{figure}
\centering
\includegraphics[width=0.50\textwidth]{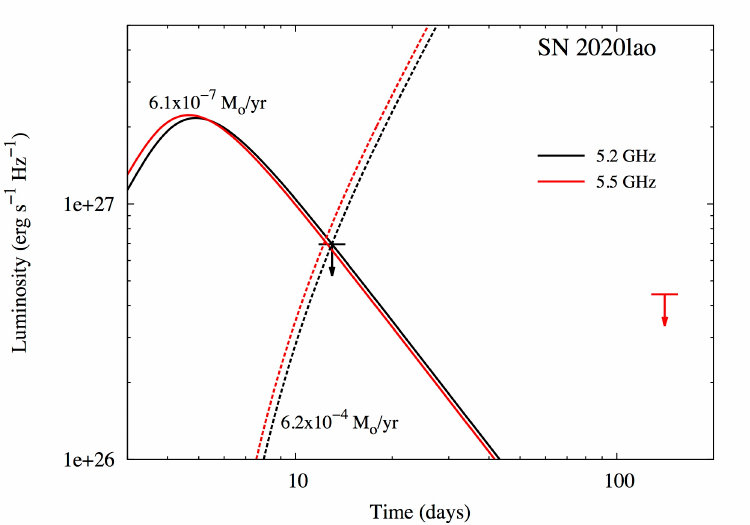} 
\caption{Observed upper limits at 5.2 GHz (+13 days, black) and at 5.5 GHz (+141 days, red), together with modeled light curves for two mass-loss rates: $\dot{M} = 6.1\times10^{-7}\,M_{\odot}\,\mathrm{yr}^{-1}$ (solid lines) and $\dot{M} = 6.2\times10^{-4}\,M_{\odot}\,\mathrm{yr}^{-1}$ (dashed lines).}
\label{fig:radio2020lao}
\end{figure}

To interpret these radio limits in the context of the explosion dynamics, we computed synchrotron models \citep[see][for details of the modeling]{Harris23,Venkattu24} based on explosion parameters inferred from the bolometric light curve ($E_k \simeq \EKnoerr \times10^{51}$\,erg, $M_{\mathrm{ej}} \simeq 3.2\,M_{\odot}$) an outer density profile characterized by a power-law index of $n=10$, and a wind density profile ($\rho \propto r^{-2}$). For these parameters, the transition velocity separating the flat inner ejecta from the steep outer component occurs at $\approx 2.94\times10^{4}$\ \kms. For a nominal WR-like wind with mass-loss rate $\dot{M} = 6.1\times10^{-7}\,M_{\odot}\,\mathrm{yr}^{-1}$ and wind velocity $v_{\mathrm{w}} = 1000$\ \kms\ the shock propagates at roughly $0.57c$ at +13 days, substantially faster than the canonical $10^{5}$~\kms\ often assumed for ordinary SE~SNe. This high shock velocity places SN~2020lao in the regime illustrated in Fig.~6 of \citet{ChevalierFransson2017}, where radio emission rises and peaks on very short timescales and is entirely optically thin for $t \gtrsim 7-8$\,days at 5 GHz (cf. Fig. \ref{fig:radio2020lao}). In comparison, SN~1998bw reached $L_{\nu} \sim 10^{29}$\,erg\,s$^{-1}$\,Hz$^{-1}$ at $\sim$11 days with shock speeds $\gtrsim c/3$.

Our synchrotron models adopt a WR-like composition for the wind (50\% C and 50\% O by mass), which yields slightly higher electron densities and marginally more efficient radio emission than cosmic abundances. The modeled 5.2 GHz and 5.5 GHz luminosities are shown in Fig.~\ref{fig:radio2020lao}. In these models, we have also included free-free absorption in the unshocked, yet fully ionized, C and O-rich wind, assuming a wind temperature of $3\times10^4$ K \citep[see][]{Lundqvist88}.
Adopting $\dot{M} = 6.1\times10^{-7}\,M_{\odot}\,\mathrm{yr}^{-1}$ and $v_{\mathrm{w}} = 1000$\,km\,s$^{-1}$, the model predicts a peak (due to synchrotron self-absorption) at $t_{\mathrm{pk}} \simeq 5$\,days with a 5.2\,GHz luminosity of $\simeq 2.2\times10^{27}$\,erg\,s$^{-1}$\,Hz$^{-1}$. This is consistent with expectations from Fig.~6 of \citet{ChevalierFransson2017}. The model also matches the observed 13-day limit of $L_{\nu} \lesssim 6.9\times10^{26}$\,erg\,s$^{-1}$\,Hz$^{-1}$ at 5.2 GHz.
A caveat is that the high shock velocity at this epoch ($\approx 0.57c$) makes our nonrelativistic approach only marginally valid, but the qualitative conclusion remains robust: SN~2020lao would have produced a detectable radio signal at 13\,days unless the external density were exceptionally low.

A complication is that for high mass-loss rates, free–free absorption from the wind starts to suppress early radio emission, so that the 13-day limit cannot rule out mass-loss rates $\dot{M}\gtrsim6.2\times10^{-4}\,M_{\odot}\,\mathrm{yr}^{-1}$  
(see Fig.\,\ref{fig:radio2020lao}). However, the late-time (141\,day) limit at 5.5 GHz breaks this degeneracy: for $\dot{M} = 6.2\times10^{-4}\,M_{\odot}\,\mathrm{yr}^{-1}$ the model predicts a 5.5\,GHz luminosity of $L_{\nu} \approx 1.3\times10^{29}$\,erg\,s$^{-1}$\,Hz$^{-1}$ at 141\,days, which is firmly excluded. Combining the two epochs therefore yields the constraint

\begin{equation}
\dot{M} \lesssim 6\times10^{-7}\,
\left( \frac{v_{\mathrm{w}}}{1000\,\mathrm{km\,s^{-1}}} \right)
\,M_{\odot}\,\mathrm{yr}^{-1}.
\end{equation}

\begin{figure}
     \centering
     \includegraphics[width=\linewidth]{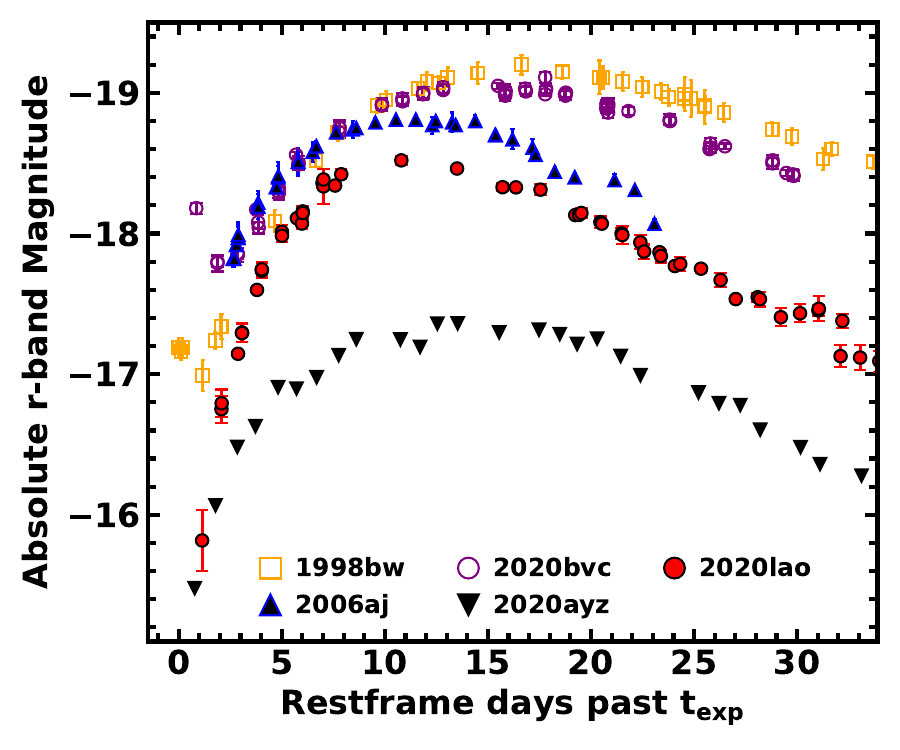}
     \caption{Absolute $r$-band magnitude light curve of SN~2020lao compared with several well-studied SNe~Ic~BL. These include the $R/r$-band light curves of GRB980425/SN~1998bw ($A^{MW}_{R} = 0.12$ mag, $\mu = 32.9$ mag; \citealt{Clocchiatti2011}), XRF~060218/SN~2006aj ($A^{tot}_R = 0.04$ mag, $\mu = 35.7$ mag; \citealt{Ferrero2006}), and SN~2020bvc ($A^{MW}_r = 0.08$ mag, $\mu = 35.4$ mag), which also exhibited prominent early excess emission and high-energy signatures possibly produced from cocoon emission \citep{Ho2020,Izzo2020}. The ZTF $r$-band light curve of SN~2020ayz ($A^{MW}_r = 0.03$ mag, $\mu = 35.8$ mag) is also included for comparison; like SN~2020lao, it was discovered shortly after explosion and lacks signatures of early excess emission \citep{Ho2020}.}
\label{fig:rband_comp}
 \end{figure}

\textit{Swift} X-Ray Telescope  (XRT) observations at $\sim$14~days yielded no detection, with a flux limit of $F_X \lesssim 2.9\times10^{-14}$~erg~s$^{-1}$~cm$^{-2}$ ($L_X \lesssim 6\times10^{40}$~erg~s$^{-1}$ at $D=131.8$~Mpc; \citealt{Corsi2023}).
This is consistent with expectations for ordinary core-collapse SNe \citep{Dwarkadas2025} and significantly fainter than the afterglows of classical, cosmological GRBs \citep{Kann2011}.
Interpreted within the inverse-Compton framework for SE SNe interacting with a wind \citep[e.g.,][]{Chevalier2006, Margutti2012}, the X-ray non-detection implies a very low external density, consistent with the WR-like environment inferred from the radio limits, and supports the conclusion that SN~2020lao showed no detectable relativistic outflow or central-engine powered emission associated with canonical events such as SN~1998bw/GRB~980425 and SN~2006aj/XRF~060218.

More recently, fast X-ray transient SNe~Ic BL identified by the \textit{Einstein} Probe (EP) provide an additional comparison class for interpreting the multiwavelength limits of SN~2020lao. For example, the well-studied EP~250108a/SN~2025kg \citep{EylesFerris2025,Rastinejad2025,Srinivasaragavan2025,Li2025} differs from SN~2020lao in exhibiting early optical excess emission and post-maximum spectroscopic signatures of circumstellar interaction. Despite these differences, and despite being identified through prompt X-ray emission, EP~250108a/SN~2025kg was not detected by \textit{Swift}/XRT at approximately two days post explosion and showed no radio emission at any epoch, with MeerKAT 3.06\,GHz limits at the level of about 24\,$\mu$Jy (3$\sigma$), corresponding to a luminosity limit of $L_{\nu} \lesssim 1.8\times10^{28}$\,erg\,s$^{-1}$\,Hz$^{-1}$ at $z=0.176$ \citep{Srinivasaragavan2025}. The \textit{Swift}/XRT upper limit for SN~2020lao at about 14 days post explosion, together with the deep radio non-detections at 13 and 141 days, are comparable in flux density to those reported for EP~250108a/AT~2025kg, while being more constraining in luminosity because of the much smaller distance, reaching $L_{\nu} \lesssim 10^{27}$\,erg\,s$^{-1}$\,Hz$^{-1}$. These comparisons show that although SN~2020lao lacked a luminous, long-lived afterglow characteristic of classical GRBs, its X-ray and radio limits overlap the regime occupied by weak or rapidly fading high-energy transients associated with some low-luminosity GRBs and fast X-ray transient SNe~Ic BL, which may be powered by a combination of magnetar-driven energy injection and radioactive heating \citep[e.g.,][]{Zhu2025,Srinivasaragavan2025b}.

\subsection{SN~2020lao in comparison with other SNe~Ic-BL}

Placing SN~2020lao in context with other SNe~Ic-BL helps clarify its explosion properties and highlight where it fits within the broader parameter space of the class. Although the full set of comparison spectra is not shown here, we examined SN~2020lao alongside a wide sample of well-observed SNe~Ic-BL. From these comparisons, we find that SN~2020lao most closely resembles SN~2006aj in its overall spectral morphology, but with markedly higher Doppler velocities. Inspection of  Fig.~\ref{fig:specseq} clearly reveals that the line profiles of the two events are similar in form, yet the absorption minima in SN~2020lao are significantly more blueshifted, indicating much more rapidly expanding ejecta. Consistent with this, the spectral-synthesis analysis in Sect.~\ref{sec:spectralsynthesis} demonstrates that SN~2020lao requires substantially larger values of $E_K$ and $E_K/M_{ej}$ than SN~2006aj.

Figure~\ref{fig:rband_comp} compares the absolute $r$-band light curve of SN~2020lao with those of SNe~Ic-BL both with and without signatures of engine-driven activity. The sample includes events associated with high-energy emission or early optical excesses often interpreted as afterglow or cocoon emission (SNe~1998bw, 2006aj,\footnote{SN~2006aj exhibits early excess emission primarily in bluer bands \citep[][and references therein]{modjaz2019}.} and 2020bvc), as well as SN~2020ayz, a rapidly discovered SN~Ic-BL lacking such features. SN~2020lao shows no clear early excess, and its light-curve evolution is broadly consistent with that of other SNe~Ic-BL, regardless of whether they exhibit early high-energy signatures.

For additional context, we compared SN~2020lao with the canonical GRB-associated SN~1998bw. Spectral-synthesis modeling of SN~1998bw, assuming spherical symmetry, yields $M_{ej}\approx 8$--$10\,M_\odot$ and $E_K\approx (30$--$50)\times10^{51}$~erg, corresponding to $E_K/M_{ej}$  $\sim(3$--$6)\times10^{51}$~erg~$M_\odot^{-1}$. Although SN~1998bw was likely highly aspherical, whereas SN~2006aj exhibited much weaker asymmetry, these values nonetheless provide a useful benchmark. We note that this comparison refers to the bulk SN ejecta and should be distinguished from the isotropic-equivalent prompt gamma-ray energy of GRB~980425 ($E_{\rm iso}\sim 8\times10^{47}$~erg), which is orders of magnitude smaller and traces a physically distinct component of the explosion.

When compared to the $E_K/M_{ej}$ distribution of SNe~Ic-BL, SN~2020lao lies near the upper end, approaching the energetics of SN~1998bw while more closely resembling SN~2006aj in spectral morphology. Yet despite its high kinetic energy, it shows none of the usual signatures of engine-driven explosions: its early optical light curve lacks any excess attributable to afterglow or cocoon emission, and deep radio and X-ray limits rule out luminous relativistic ejecta. Taken together, these constraints indicate a low-density, WR-like environment and are strong evidence against a detectable cocoon breakout or relativistic afterglow emission components.
Despite its unusually high specific kinetic energy, SN~2020lao falls among the radio- and X-ray-quiet SNe~Ic-BL rather than the engine-driven events exemplified by SN~1998bw/GRB~980425. If a central engine was active and launched a jet, the observations require that it was either viewed far off-axis or choked before breakout. Alternatively, SN~2020lao may represent an extreme but nonrelativistic SN~Ic-BL lacking sustained engine activity, though this would be somewhat surprising given its large high spectral line velocities.

\section{Conclusion}

We  have presented an exceptional  early-time optical dataset of the SN~Ic-BL~2020lao, including optical photometry beginning within hours of the explosion along with a high-cadence spectroscopic time series beginning within $\approx$48 hours  of the explosion and extending to nearly one hundred days later. The explosion epoch is well constrained thanks to continuous TESS coverage and complementary limits and detections from the ZTF and ATLAS  public data streams.  

The light curves of SN~2020lao show a rapid rise of $\approx 10$ days and a peak absolute $r$-band magnitude of $M_r \simeq -18.5$, typical of SNe~Ic-BL \citep[e.g.,][]{taddia2019}. Its spectra resemble those of SN~2006aj but with systematically higher expansion velocities. Line-velocity measurements and Arnett-model fits, combined with an appropriate $E_K/M_{ej}$ relation, yield a $^{56}$Ni mass of $\sim \Nimass M_{\odot}$, $M_{ej} = \Mej\ M_{\odot}$, $E_K \sim \EK \times 10^{51}$~erg, and $E_K/M_{ej} = \EKoverM \times 10^{51}$~erg~$M_{\odot}^{-1}$. 
Spectral synthesis with a scaled SN~2006aj density structure reproduces the overall photospheric-phase spectra and is consistent with the explosion parameters inferred from the light curve:$M_{ej} = \Mejsyn\ M_{\odot}$, $E_K \approx \EKsyn \times 10^{51}$~erg, and $E_K/M_{ej} \approx \EoverMsyn \times 10^{51}$~erg $M_{\odot}^{-1}$.

Compared with analogous parameter estimates for SN~2006aj derived from similar spectral-synthesis modeling \citep{Mazzali2006}, SN~2020lao shows a comparable $^{56}$Ni mass of $\sim 0.2\,M_{\odot}$ but an $E_K/M_{ej}$ ratio that is a factor of $\gtrsim 5$ higher. Moreover, the $E_K/M_{ej}$ values inferred for SN~2020lao are higher than those of SN~Ic-BL events such as SNe~2006aj and 2020bvc, aligning it more closely with the energetics of GRB-associated SNe~Ic-BL such as SN~1998bw. 

All this taken together, SN~2020lao exemplifies an energetic SN~Ic-BL, a property often associated with engine-driven explosions. Yet its early optical behavior, combined with deep radio and X-ray non-detections, provides no evidence of any of the afterglow emission that would be expected from such activity. This combination implies that, if a jet was launched, the event must have been viewed far off-axis or the jet was choked before breakout. An alternative possibility is that SN~2020lao represents an extreme but purely nonrelativistic SN~Ic-BL without sustained central-engine activity. Systematic infant-phase follow-up, including sensitive radio to X-ray monitoring, polarimetry, and nebular spectroscopy to probe the ejecta geometry, will be essential for assessing how common such engine-quiet yet energetic SNe Ic-BL are and for distinguishing viewing-angle effects from intrinsic physical differences. In conclusion, SN 2020lao establishes a valuable empirical baseline that we hope will aid future efforts to separate the underlying SN emission in SNe Ic-BL from the additional components powered by central-engine activity in GRB- and XRF-associated events such as SN 2006aj.

\section*{Data availability}

The spectra presented in this work are publicly available through the WISeREP repository \citep{Ofer2012} and can be downloaded from \url{https://www.wiserep.org/object/14810}. Broadband photometry of SN~2020lao are available in electronic form at the CDS via anonymous ftp to cdsarc.u-strasbg.fr (130.79.128.5) or via \url{http://cdsweb.u-strasbg.fr/cgi-bin/qcat?J/A+A/}.

\begin{acknowledgements}

We thank Morgan Fraser, Francisco Förster, Rasmus Handberg, and Simon Holmbo for insightful discussions and/or providing observational support to the NUTS2 programme. 
Support for the Aarhus supernova group comes from Independent Research Fund Denmark (IRFD; grant numbers 8021-00170B and 10.46540/2032-00022B), and  by the Aarhus University Research Foundation Nova project (AUFF-E-2023-9-28).
TJM is supported by the Grants-in-Aid for Scientific Research of the Japan Society for the Promotion of Science (JP24K00682, JP24H01824, JP21H04997, JP24H00002, JP24H00027, JP24K00668) and by the Australian Research Council (ARC) through the ARC's Discovery Projects funding scheme (project DP240101786).
L.G. acknowledges financial support from AGAUR, CSIC, MCIN and AEI 10.13039/501100011033 under projects PID2023-151307NB-I00, PIE 20215AT016, CEX2020-001058-M, ILINK23001, COOPB2304, and 2021-SGR-01270.
NER, AP and GV acknowledge support from the PRIN-INAF 2022 “Shedding light on the nature of gap transients: from the observations to the models”.
EP acknowledges financial support from INAF and kind hospitality at NAOJ, Mitaka, Tokyo.
SM is funded by Leverhulme Trust grant RPG-2023-240. \\

\indent Partial funding enabling NUTS2 use of the Nordic Optical Telescope (NOT) comes from the Instrument Center for Danish Astrophysics (IDA). The NOT is owned in collaboration by the University of Turku and Aarhus University, and operated jointly by Aarhus University, the University of Turku and the University of Oslo, representing Denmark, Finland and Norway, and the University of Iceland at the Observatorio del Roque de los Muchachos, La Palma, Spain, of the Instituto de Astrofisica de Canarias.
NOT observations were obtained under program ID P61-507 and P62-505. 
This work makes use of observations from the Las Cumbres Observatory Global  Telescope network, and also from the GTC under programmes GTC52-20A and GTC69-20A.
This research also makes use of Astropy, a community-developed core Python package for Astronomy \citep{2018AJ....156..123A}, as well as Photutils \citep{larry_bradley_2020_4044744}. 

\end{acknowledgements}

 \bibliographystyle{aa}
\bibliography{bibtex.bib}

\FloatBarrier
\begin{appendix}

\onecolumn 
\section{Spectroscopic journal}

\normalsize 
\setlength{\tabcolsep}{4pt}

\begin{longtable}{llllllr}
\caption{\normalsize  Journal of spectroscopic observations.\label{tab:specjor}}\\
\toprule
Date & JD\,2,400,000 & Phase$^{a}$ & Telescope$^{b}$ & Instrument & Grating & Exp.\ (s) \\
\midrule
\endfirsthead

\caption[]{Journal of Spectroscopic Observations (continued).}\\
\toprule
Date & JD\,2,400,000 & Phase$^{a}$ & Telescope$^{b}$ & Instrument & Grating & Exp.\ (s) \\
\midrule
\endhead

\midrule
\multicolumn{7}{r}{Continued on next page}\\
\endfoot

\bottomrule
\endlastfoot

\multicolumn{7}{c}{Optical}\\
\midrule
\rowcolor{gray!15}
2020 May 26.2 & 58995.69 & +2.0 & GTC$^{c}$ & OSIRIS & R1000B,R & 800\\
2020 May 27.1 & 58996.64 & +2.9 & GTC & OSIRIS & R1000B,R & 700\\
\rowcolor{gray!15}
2020 May 27.9 & 58996.99 & +3.7 & NOT & ALFOSC & Gr4 & 2100\\
2020 May 29.5 & 58999.00 & +5.2 & Faulkes & Floyds & Red, Blue & 2400\\
\rowcolor{gray!15}
2020 May 30.2 & 58999.65 & +5.8 & NOT & ALFOSC & Gr4 & 2100\\
2020 May 31.0 & 59000.46 & +6.6 & LT & Sprat & VPH & 500\\
\rowcolor{gray!15}
2020 June 1.5 & 59002.01 & +8.1 & Faulkes$^{d}$ & Floyds & Red, Blue & 3600\\
2020 June 5.3 & 59005.80 & +11.8 & APO & DIS & Red, Blue & 2100\\
\rowcolor{gray!15}
2020 June 6.9 & 59007.44 & +13.4 & NOT & ALFOSC & Gr4 & 2100\\
2020 June 19.0 & 59019.90 & +25.0 & LT & Sprat & VPH & 1200\\
\rowcolor{gray!15}
2020 June 21.0 & 59021.54 & +27.0 & NOT & ALFOSC & Gr4 & 900\\
2020 July 6.0 & 59036.51 & +41.5 & NOT & ALFOSC & Gr4 & 3600\\
\rowcolor{gray!15}
2020 August 4.0 & 59067.50 & +71.5 & GTC & OSIRIS & R1000B & 2$\times$1500\\
2020 August 31.1 & 59092.69 & +95.9 & GTC & OSIRIS & R1000B,R & 785\\

\midrule
\multicolumn{7}{c}{NIR}\\
\midrule
\rowcolor{gray!15}
2020 June 15.4 & 59015.87 & +21.5 & IRTF & SpeX & Prism & 1357\\
\end{longtable}
\tablefoot{\\ \small 
\tablefoottext{a}{Rest-frame days past $t_{exp}$.}
\tablefoottext{b}{GTC is the 10.4-m Gran Telescopio Canarias, NOT is the 2.56-m Nordic Optical Telescope, Faulkes is the Las Cumbres Observatory's 2-m  telescope, LT is the 2-m Liverpool Telescope, APO is the 3.5-m  Apache Point Observatory telescope, and IRTF is the 3.2-m NASA Infrared Telescope Facility.}
\tablefoottext{c}{Spectrum   as reported by \citet{Galbany2025}.}
\tablefoottext{d}{Classification spectrum from the \href{https://www.wis-tns.org/object/2020lao}{Transient Name Server}  as reported by \citet{Burke2020}.}
}

\end{appendix}

\end{document}